%% file: houch04.tex
\input harvmac 
\input epsf.tex

\overfullrule=0mm
\def\file#1{#1}
\def\figbox#1#2{\epsfxsize=#1\vcenter{
\epsfbox{\file{#2}}}} 
\newcount\figno
\figno=0
\def\fig#1#2#3{
\par\begingroup\parindent=0pt\leftskip=1cm\rightskip=1cm\parindent=0pt
\baselineskip=11pt
\global\advance\figno by 1
\midinsert
\epsfxsize=#3
\centerline{\epsfbox{#2}}
\vskip 12pt
{\bf Fig.\the\figno:} #1\par
\endinsert\endgroup\par
}
\def\figlabel#1{\xdef#1{\the\figno}}
\def\encadremath#1{\vbox{\hrule\hbox{\vrule\kern8pt\vbox{\kern8pt
\hbox{$\displaystyle #1$}\kern8pt}
\kern8pt\vrule}\hrule}}


\def\IR{\relax{\rm I\kern-.18em R}}
\font\cmss=cmss10 \font\cmsss=cmss10 at 7pt

\font\cmss=cmss10 \font\cmsss=cmss10 at 7pt
\def\IZ{\relax\ifmmode\mathchoice
{\hbox{\cmss Z\kern-.4em Z}}{\hbox{\cmss Z\kern-.4em Z}}
{\lower.9pt\hbox{\cmsss Z\kern-.4em Z}}
{\lower1.2pt\hbox{\cmsss Z\kern-.4em Z}}\else{\cmss Z\kern-.4em Z}\fi}
\def\IN{\relax{\rm I\kern-.18em N}}
\def\Rho{{\bf r}}
\def\b{\circ}
\def\n{\bullet}

\def\gbbbb{\Gamma_4^{\hbox{$\scriptstyle \b \b$}\kern -8.2pt
\raise -4pt \hbox{$\scriptstyle \b \b$}}}
\def\gnnnn{\Gamma_4^{\hbox{$\scriptstyle \n \n$}\kern -8.2pt  
\raise -4pt \hbox{$\scriptstyle \n \n$}}}
\def\gnnnnnn{\Gamma_6^{\hbox{$\scriptstyle \n \n \n$}\kern -12.3pt
\raise -4pt \hbox{$\scriptstyle \n \n \n$}}}
\def\gbbbbbb{\Gamma_6^{\hbox{$\scriptstyle \b \b \b$}\kern -12.3pt
\raise -4pt \hbox{$\scriptstyle \b \b \b$}}}
\def\gbbbbc{\Gamma_{4\, c}^{\hbox{$\scriptstyle \b \b$}\kern -8.2pt
\raise -4pt \hbox{$\scriptstyle \b \b$}}}
\def\gnnnnc{\Gamma_{4\, c}^{\hbox{$\scriptstyle \n \n$}\kern -8.2pt
\raise -4pt \hbox{$\scriptstyle \n \n$}}}
\def\Rbud#1{{\cal R}_{#1}^{-\kern-1.5pt\blacktriangleright}}
\def\Rleaf#1{{\cal R}_{#1}^{-\kern-1.5pt\vartriangleright}}
\def\Rbudb#1{{\cal R}_{#1}^{\circ\kern-1.5pt-\kern-1.5pt\blacktriangleright}}
\def\Rleafb#1{{\cal R}_{#1}^{\circ\kern-1.5pt-\kern-1.5pt\vartriangleright}}
\def\Rbudn#1{{\cal R}_{#1}^{\bullet\kern-1.5pt-\kern-1.5pt\blacktriangleright}}
\def\Rleafn#1{{\cal R}_{#1}^{\bullet\kern-1.5pt-\kern-1.5pt\vartriangleright}}
\def\Wleaf#1{{\cal W}_{#1}^{-\kern-1.5pt\vartriangleright}}
\def\Cleaf{{\cal C}^{-\kern-1.5pt\vartriangleright}}
\def\Cbud{{\cal C}^{-\kern-1.5pt\blacktriangleright}}
\def\Crleaf{{\cal C}^{-\kern-1.5pt\circledR}}

\writetoc


\Title{\vbox{\hsize=3.truecm \hbox{SPhT/04/078}}}
{{\vbox {
\bigskip
\centerline{2D Quantum Gravity, Matrix Models}
\medskip
\centerline{and Graph Combinatorics}
}}}
\bigskip
\centerline{P. Di Francesco\foot{philippe@spht.saclay.cea.fr}}
\medskip
\centerline{ \it Service de Physique Th\'eorique, CEA/DSM/SPhT}
\centerline{ \it Unit\'e de recherche associ\'ee au CNRS}
\centerline{ \it CEA/Saclay}
\centerline{ \it 91191 Gif sur Yvette Cedex, France}
\bigskip
\noindent  Lectures given at the summer school 
``Applications of random matrices in physics", Les Houches, June 2004. 

\Date{06/04}
\nref\CFT{ P. Di Francesco, P. Mathieu and D. S\'en\'echal,
{\it Conformal Field Theory}, Graduate Texts in Contemporary Physics,
Springer (1996).}
\nref\KPZ{V.G. Knizhnik, A.M. Polyakov and A.B. Zamolodchikov, Mod. Phys. Lett.
{\bf A3} (1988) 819; F. David, Mod. Phys. Lett. {\bf A3} (1988) 1651; J.
Distler and H. Kawai, Nucl. Phys. {\bf B321} (1989) 509.}
\nref\TUT{W. Tutte,
{\it A Census of Planar Maps}, Canad. Jour. of Math. 
{\bf 15} (1963) 249-271; {\it A Census of planar triangulations},
Canad. Jour. of Math. {\bf 14} (1962) 21-38; 
{\it A Census of Hamiltonian polygons}, 
Canad. Jour. of Math. {\bf 14} (1962) 402-417.}
\nref\NOB{G. 't Hooft, Nucl. Phys. {\bf B72} (1974) 461}
\nref\BIPZ{E. Br\'ezin, C. Itzykson, G. Parisi and J.-B. Zuber, {\it Planar
Diagrams}, Comm. Math. Phys. {\bf 59} (1978) 35-51.}
\nref\SURF{V. Kazakov, Phys. Lett. {\bf 150B} (1985) 282; F. David, Nucl.
Phys. {\bf B257} (1985) 45; J. Ambjorn, B. Durhuus and J. Frohlich,
Nucl. Phys. {\bf B257} (1985) 433; V. Kazakov, I. Kostov and A. Migdal,
Phys. Lett. {\bf 157B} (1985) 295.}
\nref\THREESOME{E. Br\'ezin and V. Kazakov, {\it Exactly solvable field
theories of closed strings}, Phys. Lett. {\bf B236} 
(1990) 144-150; M. Douglas and S. Shenker, {\it Strings in less than 1 dimension},
Nucl. Phys. {\bf B335} (1990) 635;
D. Gross and A. Migdal, {\it Non-perturbative two-dimensional gravity},
Phys. Rev. Lett. {\bf 64} (1990) 127-130.}
\nref\GRAVI{see for instance the review  by P. Di Francesco, 
P. Ginsparg and J. Zinn-Justin,
{\it 2D Gravity and Random Matrices}, 
Physics Reports {\bf 254} (1995) 1-131.}
\nref\WIKO{E. Witten, {\it On the topological phase of two-dimensional gravity},
Nucl. Phys. {\bf B340} (1990) 281-332 and  
{\it Two dimensional gravity and intersection theory on moduli space},
Surv. in Diff. Geom. {\bf 1} (1991) 243-310; M. Kontsevich, {\it Intersection 
theory on the moduli space of curves and the matrix Airy function},
Comm. Math. Phys. {\bf 147} (1992) 1-23.}
\nref\BECAN{E. Bender and E. Canfield, {\it The number of degree-restricted 
rooted maps on the sphere}, SIAM J. Discrete Math. {\bf 7}(1) (1994) 9-15.}
\nref\SCHth{G. Schaeffer, {\it Conjugaison d'arbres
et cartes combinatoires al\'eatoires} PhD Thesis, Universit\'e 
Bordeaux I (1998) and {\it Bijective census and random
generation of Eulerian planar maps}, Electronic
Journal of Combinatorics, vol. {\bf 4} (1997) R20.}
\nref\CENSUS{J. Bouttier, P. Di Francesco and E. Guitter, {\it Census of planar
maps: from the one-matrix model solution to a combinatorial proof},
Nucl. Phys. {\bf B645}[PM] (2002) 477-499, arXiv:cond-mat/0207682.}
\nref\CS{R. Cori, {\it Un code pour les graphes planaires et ses
applications}, Soci\'et\'e Math\'ematique de France, Paris 1975; D. Arqu\`es,
{\it Les hypercartes planaires sont des arbres tr\`es bien \'etiquet\'es},
Discr. Math. {\bf 58}(1) (1986) 11-24; M. Marcus and G. Schaeffer, 
{\it Une bijection simple pour les cartes orientables}, 
http://www.loria.fr/$\sim$schaeffe/pub/orientable/directe.ps; 
P. Chassaing and G. Schaeffer, {\it Random Planar Lattices and 
Integrated SuperBrownian Excursion}, to appear in 
Probability Theory and Related Fields, arXiv:math.CO/0205226.}
\nref\GEOD{J. Bouttier, P. Di Francesco and E. Guitter, {\it Geodesic
distance in planar graphs}, Nucl. Phys. B 663[FS] (2003) 535-567, 
arXiv:cond-mat/0303272.}
\nref\STAU{M. Staudacher, {\it The Yang-Lee Edge Singularity on a
Dynamical Planar Random Surface}, Nucl. Phys. {\bf B336} (1990) 349-362.}
\nref\ON{ I. Kostov, Mod. Phys. Lett. {\bf A4} (1989) 217; M. Gaudin and
I. Kostov, Phys. Lett. {\bf B220} (1989) 200; I. Kostov and M. Staudacher, Nucl.
Phys. {\bf B384} (1992) 459;B. Eynard and J. Zinn-Justin, Nucl. Phys. {\bf B386} (1992) 558;
B. Eynard and C. Kristjansen, Nucl. Phys. {\bf B455} (1995) 577 and Nucl. Phys. 
{\bf B466} (1996) 463-487.}
\nref\QPO{D. Boulatov and V. Kazakov, {\it The Ising model
on a random planar lattice: the structure of the phase 
transition and the exact critical exponents}, Phys. Lett. {\bf B186} (1987) 379-384; 
V. Kazakov, Nucl. Phys. {\bf B4} (Proc. Suppl.) (1998), 93; J.-M. Daul,
{\it Q-states Potts model on a random planar lattice},
arXiv:hep-th/9502014; 
B. Eynard and G. Bonnet, {\it The Potts-q random matrix model:
loop equations, critical exponents, and rational case}, Phys. Lett. {\bf B 463}
(1999) 273-279, arXiv:cond-mat/9906130;
P. Zinn-Justin, {\it The dilute Potts model on random
surfaces}, J. Stat. Phys. {\bf 98} (2001) 245-264, arXiv:cond-mat/9903385.}
\nref\IRFK{I. Kostov,{\it Strings with discrete target space}, Nucl. Phys. {\bf B376}
(1992) 539-598; {\it Gauge invariant matrix models for A-D-E closed strings}, 
Phys. Lett. {\bf B297} (1992) 74-81.}
\nref\RECT{P. Di Francesco, {\it Rectangular Matrix Models and Combinatorics
of Colored Graphs}, Nucl. Phys. {\bf B648} (2002) 461-496, arXiv:cond-mat/0207682.}
\nref\ORTHO{D. Bessis, Comm. Math. Phys. {\bf 69} (1979) 147-163;
D. Bessis, C.Itzykson and J.-B. Zuber, Adv. in Appl.
Math. {\bf 1} (1980) 109.}
\nref\DK{P. Di Francesco and D. Kutasov, {\it Unitary minimal  
models coupled to gravity}, Nucl. Phys. {\bf B342} (1990) 589   and
{\it Integrable models of 2D quantum gravity} NATO ASI Series {\bf B315}
(1993) 73.} 
\nref\DS{V. Drinfeld and V. Sokolov, J. Sov. Math. {\bf 30} (1985) 1975
and Sov. Math. Dokl. {\bf 23} No.3 (1981) 457.}
\nref\BMS{M. Bousquet-M\'elou and G. Schaeffer,{\it The degree distribution
in bipartite planar maps: application to the Ising model},
arXiv:math.CO/0211070.}
\nref\NOUSHARD{J. Bouttier, P. Di Francesco and E. Guitter, {\it
Combinatorics of hard particles on planar maps}, Nucl. Phys. 
{\bf B655} (2002) 313-341, arXiv:cond-mat/0211168.}
\nref\AW{J. Ambj\o rn and Y. Watabiki, {\it Scaling in quantum gravity},
Nucl.Phys. {\bf B445} (1995) 129-144.}
\nref\ONEWALL{J. Bouttier, P. Di Francesco and E. Guitter, {\it Statistics
of planar maps viewed from a vertex: a study via labeled trees},
Nucl. Phys. {\bf B675}[FS] (2003) 631-660, arXiv:cond-mat/0307606.}
\nref\TWOWALL{J. Bouttier, P. Di Francesco and E. Guitter, {\it Random
trees between two walls: Exact partition function}, J. Phys. A: Math. Gen. 
{\bf 36} (2003) 12349-12366, arXiv:cond-mat/0306602.} 
\nref\MOB{J. Bouttier, P. Di Francesco and E. Guitter, {\it Planar maps as labeled mobiles},
arXiv:math.CO/0405099 (2004), submitted to Elec. Jour. of Combinatorics.}
\nref\ISE{D.J. Aldous, {\it Tree-based models for random distribution of mass}, J. Stat. Phys. {\bf 73}
(1993) 625-641; E. Derbez and G. Slade, {\it The scaling limit of lattice trees in high dimensions},
Comm. Math. Phys. {\bf 193} (1998) 69-104; J.-F. Le Gall, {\it Spatial branching processes,
random snakes and partial differential equations}, Lectures in Mathematics ETH Zurich, Birkh\"auser,
Basel, 1999.}
\nref\CHASDUR{P. Chassaing and B. Durhuus, {\it Statistical Hausdorff 
dimension of labelled trees and quadrangulations}, 
arXiv:math.PR/0311532.}
\nref\MARMO{J. F. Marckert and A. Mokkadem, {\it Limit of normalized 
quadrangulations: the Brownian map}, arXiv:math.PR/0403398.} 

\input toc.tex

\newsec{Introduction}

\subsec{Matrix models {\it per se}}

The purpose of these lectures is to present basic matrix models as
practical combinatorial tools, that turn out to be ``exactly solvable".
In short, a matrix model is simply a statistical ensemble of matrices with
some specific measure, here given as an invariant weight, to be integrated over
the relevant matrix ensemble. So solving a matrix model really amounts to 
computing integrals over matrix ensembles.

The lectures will be divided into two steps: first we show how to interpret 
such matrix integrals in terms of discrete two-dimensional quantum gravity, namely
in terms of graphs with prescribed topology and valences, carrying also configurations
of statistical ``matter" models;
in a second step, we show how to compute these integrals explicitly. The main difficulty
here is that the immense power of matrix integrals allows to get right and simple
answers, but gives no really good reason for such simplicity, except for technical miracles
that are sometimes called ``integrability".
To compensate for this lack of understanding,
we will always try to develop parallelly to the matrix model techniques and calculations
some purely combinatorial reading of the various results. 

The simplest combinatorial objects in many respects are trees, and we will see, at least
in the planar case, how graphs representing discrete surfaces of genus zero are reducible
to decorated trees. This eventually explains the simplicity of the corresponding
matrix model results. By pushing these ideas a little further, we will be able to 
investigate refined properties of discrete surfaces (graphs), involving their intrinsic geometry.
For instance we will compute correlation functions for surfaces with marked 
points at a prescribed geodesic distance from one-another.

Having collected many exact solutions for models of discrete geometry, it is natural to
go to the continuum limit, which displays a rich singularity structure: indeed singularities
may arise from the graphs themselves, say when parameters coupled
to valences reach some critical values, and the contribution from large graphs start dominating the
statistical sum. They may also arise from criticity of the matter statistical models defined on these
already critical graphs, in which case collective behaviors start dominating configurations.
The matrix models allow for taking both limits simultaneously (the so-called
{\it double-scaling limit}) while keeping track of all genera. 
The continuum model is expected to be described by conformally invariant
matter field theories \CFT\ coupled to 2D quantum gravity, i.e. defined on random surfaces \KPZ. 
Similarly we will write continuum correlation functions of the geodesic distance on the corresponding
random surfaces.

\subsec{A brief history}

Planar graphs first arose in combinatorics, in the groundbreaking works of Tutte \TUT\ in the 60's, 
who was able
to compute generating functions for many classes of such objects, usually called maps by combinatorists.
Higher genus was not considered then, and came up only later in physics works. 
The intrusion of matrix models in this subject occurred with the fundamental observation,
due to t'Hooft \NOB\ in the 70's, that planar graphs appearing in QCD with a large number of colors could be
viewed as Feynman diagrams for matrix models, and that moreover the size of the matrices
could serve as an expansion parameter to keep track of the topology of these diagrams. 
This caused the interest for matrix model to immediately rise, and led to the basic work
of Br\'ezin, Itzykson, Parisi and Zuber \BIPZ, who used various techniques to compute these matrix integrals,
and among other things made the contact with Tutte's enumeration results. The matrix model techniques 
were then perfected by a number of people, whose list would be too long and probably 
not exhaustive. Then came the invention of continuum and discrete quantum gravity \SURF, as the coupling
of matter theories to fluctuations of the underlying space, 
both in field-theoretical and matrix languages.
This second life of matrix models came to a climax in 1990 with three 
quasi-simultaneous papers \THREESOME\ making drastic progress in two-dimensional quantum gravity, 
as a toy model for low-dimensional non-critical strings, via the double-scaling limit of matrix models.
This started a new matrix crazyness, and certainly helped develop matrix model theory a great deal
(see \GRAVI\ for a review and references).
Remarkably, new areas of mathematics got infected by the matrix virus, thanks to Witten and Kontsevich
\WIKO,
who formulated a mathematically rigorous approach to the moduli space of punctured Riemann surfaces
using matrix models, and set the ground for a little revolution in enumerative geometry.

On the combinatorics front, it was only recently understood how to continue Tutte's work 
for higher genus
graphs or more complicated planar cases \BECAN, but a good relation to matrix model results 
is still to be found. For planar graphs however, 
the simplicity of the matrix model solutions has finally been explained 
combinatorially by Schaeffer \SCHth,
who found various bijections between planar graphs and trees, 
allowing for a simple enumeration, and a
precise contact with the matrix model solutions \CENSUS. A remarkable by-product of this approach
is that one may keep track on the trees of some features of the planar graphs, such as geodesic
distances between vertices or faces \CS\ \GEOD, a task beyond the reach of matrix models so far.

\newsec{Matrix models for 2D quantum gravity}

\subsec{Discrete 2D quantum gravity}

The purpose of quantum gravity is to incorporate in a field-theoretical setting 
the interactions between matter fields and the fluctuations of the underlying space.
In Euclidian 2D quantum gravity, the latter are represented by dynamical surfaces $\Sigma$
endowed with a Riemannian metric $g$ and scalar curvature $R$, and for which the Einstein 
action of General Relativity
reads
\eqn\generel{\eqalign{ 
S_E&=\Lambda \int_\Sigma \sqrt{g} d^2\xi  + {\cal N} \int_\Sigma \sqrt{g} R d^2\xi\cr
&=\Lambda A(\Sigma) +{\cal N} \chi(\Sigma)
\cr}}
made of a cosmological term, in which the coslological constant $\Lambda$ is coupled to the area
of the surface $A(\Sigma)$ and of the Newton term, in which the Newton constant $\cal N$
is coupled to the Euler characteristic $\chi(\Sigma)$ of the surface. The dynamical surfaces
are then discretized in the form of graphs with prescribed topology. 

We will now explain how matrix integrals can be used to generate such graphs, while
precisely keeping track of their area and their Euler characteristic. For pedagogical purposes,
we start with some simple remark on ordinary Gaussian integration, before going into
the diagrammatics of Gaussian matrix integrals.

\subsec{Gaussian integral's diagrammatics}

Consider the following Gaussian average
\eqn\gauint{ \langle x^{2n} \rangle = {1\over \sqrt{2 \pi}}
\int_{-\infty}^{\infty} e^{-{x^2 \over 2}}  x^{2n} dx  
=(2n-1)!!={(2n)!\over 2^n n!} }
Among the many ways to compute this integral, let us pick the
so-called source integral method, namely define the source
integral
\eqn\source{\Sigma(s)=\langle e^{xs}\rangle={1\over \sqrt{2\pi}} 
\int_{-\infty}^{\infty} e^{-{x^2\over 2}+sx} dx= e^{{s^2\over 2}}}
Then the average \gauint\ is obtained by taking $2n$ derivatives
of $\Sigma(s)=e^{{s^2\over 2}}$ w.r.t. $s$ and by setting $s=0$ in the end. 
It is then immediate
to see that these derivatives must be taken by {\it pairs}, in which
one derivative acts on the exponential and the other one on the prefactor
$s$. Parallelly, we note that $(2n-1)!!=(2n-1)(2n-3)...3.1$ is the 
total number of distinct combinations of $2n$ objects
into $n$ pairs. 
We may therefore formulate pictorially the computation of \gauint\ as 
follows. 
\fig{A star-diagram with one vertex and $2n$ out-coming
half-edges stands for the integrand $x^{2n}$. In the second
diagram, we have represented one non-zero contribution to $\langle
x^{2n}\rangle$ obtained by taking derivatives of $\Sigma(s)$ by
pairs represented as the corresponding pairings of half-edges
into edges.}{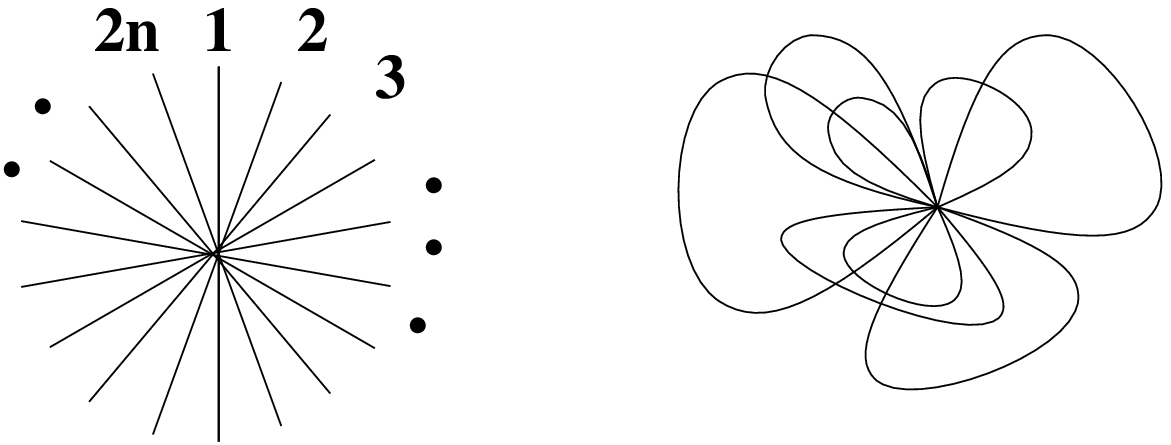}{6.cm}
\figlabel\diagone
We first draw a star-graph (see Fig.\diagone), with one central vertex and 
$2n$ outcoming
half-edges labelled $1$ to $2n$ clockwise, one for each $x$ in the integrand 
(this amounts to labelling the $x$'s in $x^{2n}$
from $1$ to $2n$). Now the pairs of derivatives
taken on the source integral are in one-to-one correspondence with pairs
of half-edges in the pictorial representation. Moreover, to get a non-zero 
contribution to
$\langle x^{2n}\rangle$, we must saturate the set of $2n$ legs by
taking $n$ pairs of them. Let us represent each such saturation
by drawing the corresponding edges as in Fig.\diagone. 
We get exactly $(2n-1)!!$
distinct labeled closed star-graphs with one vertex. This is summarized in the
one-dimensional version of Wick's theorem:
\eqn\wione{ \langle x^{2n} \rangle = \sum_{{\rm pairings}} \prod 
\langle x^2 \rangle }
where the sum extends over all pairings saturating the $2n$ half-edges,
and the weight is simply the product over all the edges thus formed
of the corresponding averages $\langle x^2\rangle =(d^2/ds^2)
\Sigma(s)\vert_{s=0}=1$.
Each saturation forms a Feynman diagram of the Gaussian average. The 
edge pairings are propagators (with value $1$ here).
This may appear like a complicated way of writing a rather trivial result,
but it suits our purposes for generalization to matrix models and graphs.

\subsec{Gaussian matrix integral and more diagrammatics}

Let us now repeat the calculations of the previous section with 
the following Gaussian Hermitian matrix average of an arbitrary
function $f$
\eqn\gaumat{ \langle f(M) \rangle = {1\over Z_0(N)} 
\int dM e^{-N Tr {M^2\over 2}}
f(M) }
where the integral extends over Hermitian $N\times N$ matrices, with the 
standard Haar measure $dM=\prod_i dM_{ii} \prod_{i<j} dRe(M_{ij}) dIm(M_{ij})$,
and the normalization factor $Z_0(N)$ is fixed by requiring that 
$\langle 1\rangle=1$ for $f=1$. Typically, we may take for $f$ a monomial
of the form $f(M)=\prod_{(i,j)\in I} M_{ij}$, $I$ a finite set of pairs
of indices. Note the presence of the normalization factor $N$ (=the size
of the matrices) in the exponential. Note that the case of the previous section 
is simply the particular case
of integration over $1\times 1$ Hermitian matrices (i.e. real numbers) here.

Like before, for a given Hermitian $N\times N$ matrix $S$, let us
introduce the source integral 
\eqn\sourmat{ \Sigma(S)=\langle e^{Tr(SM)} \rangle = e^{{Tr(S^2)\over 2N}} }
easily obtained by completing the square $M^2 -N(SM+MS)=(M-NS)^2-N^2 S^2$
and performing the change of variable $M'=M-NS$.
We can use \sourmat\ to compute any average of the form
\eqn\avform{ \langle M_{ij} M_{kl} ... \rangle={\partial \over
\partial S_{ji}} {\partial \over \partial S_{lk}} ... \ \Sigma(S)\big\vert_{S=0}}
Note the interchange of the indices due to the trace $Tr(MS)=\sum M_{ij}S_{ji}$.
As before, derivatives w.r.t. elements of $S$ must go by pairs, one
of which acts on the exponential and the other one on the $S$ element thus 
created. In particular, a fact also obvious from the parity of the Gaussian,
\avform\ vanishes unless there are an even number of matrix elements of $M$
in the average. In the simplest case of two matrix elements, we have
\eqn\prop{ \langle M_{ij} M_{kl}\rangle = {\partial\over
\partial S_{lk}} {1\over N} S_{ij} e^{{Tr(S^2)\over 2N}}\bigg\vert_{S=0}=
{1\over N}\delta_{il}\delta_{jk} }
Hence the pairs of derivatives must be taken with respect to $S_{ij}$ and
$S_{ji}$ for some pair $i,j$ of indices to yield a non-zero result. 
This leads naturally to the Matrix Wick's theorem:
\eqn\wimat{ \langle \prod_{(i,j)\in I} M_{ij} \rangle=
\sum_{\rm pairings\ P} \prod_{
(ij), (kl) \in P} \langle M_{ij} M_{kl}\rangle }
where the sum extends over all pairings saturating the (pairs of) indices
of $M$ by pairs. 

We see that in general, due to the restrictions 
\prop\ many terms in \wimat\ will vanish. Let us now give a 
pictorial interpretation for the non-vanishing contributions to \wimat.
We represent a matrix element $M_{ij}$ as a half-edge (with a marked end)
made of a double-line, each of which is oriented in an opposite direction.
We decide that the line pointing from the mark carries the index $i$,
while the other one, pointing to the mark, carries the index $j$.
This reads
\eqn\markedge{ M_{ij}\ \  \leftrightarrow \ \ \figbox{2.cm}{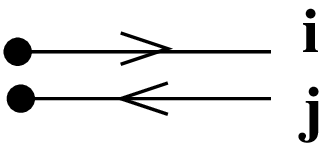} }
The two-element result \prop\ becomes simply the construction of an
edge (with both ends marked) out of two half-edges $M_{ij}$ and $M_{kl}$, 
but is non-zero 
only if the indices $i$ and $j$ are conserved along the oriented lines.
This gives pictorially
\eqn\propag{ \langle M_{ij} M_{ji} \rangle\ \   \leftrightarrow \ \ 
\figbox{3.5cm}{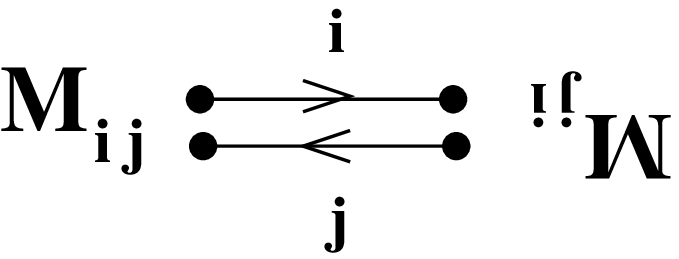} }
Similarly, an expression of the form Tr$(M^n)$ will be represented as
a star-diagram with one vertex connected to $n$ double half-edges
in such a way as to respect the identification of the various running indices,
namely
\eqn\tracrep{ {\rm Tr}(M^n)=\sum_{i_1,i_2,...,i_n} M_{i_1i_2}M_{i_2i_3}...
M_{i_ni_1} \leftrightarrow \figbox{2.5cm}{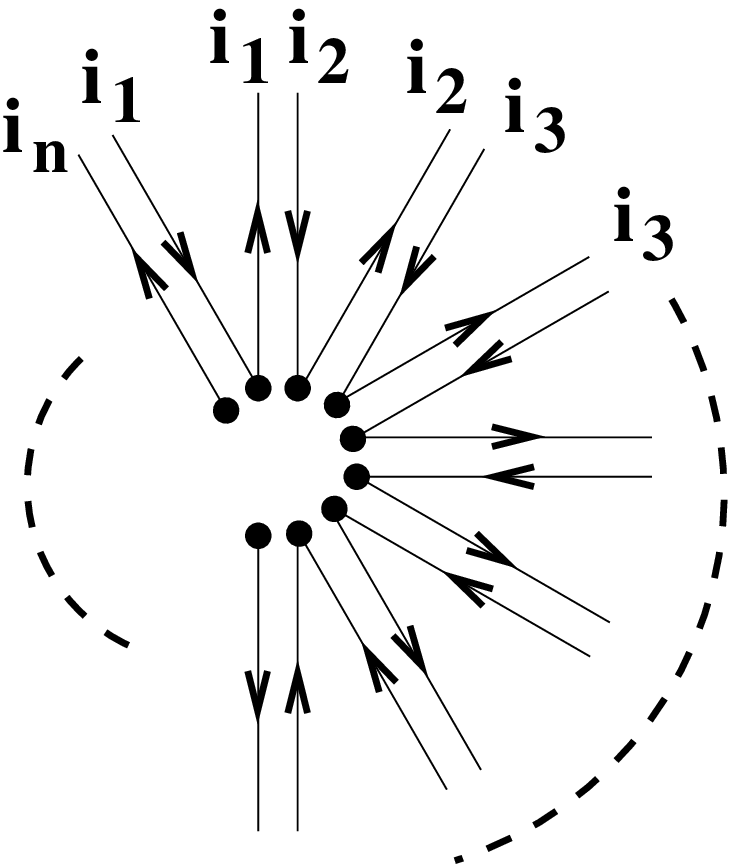} }

\fig{An example of planar (petal) diagram (a) and a non-planar one
(b). Both diagrams have $n=2p=12$ half-edges, connected with $p=6$
edges. The diagram (a) has $p+1=7$ faces bordered by oriented loops,
whereas (b) only has $3$ of them. The Euler characteristic
reads $2-2h=F-E+1$ ($V=1$ in both cases), and gives the genus $h=0$ for (a),
and $h=2$ for (b).}{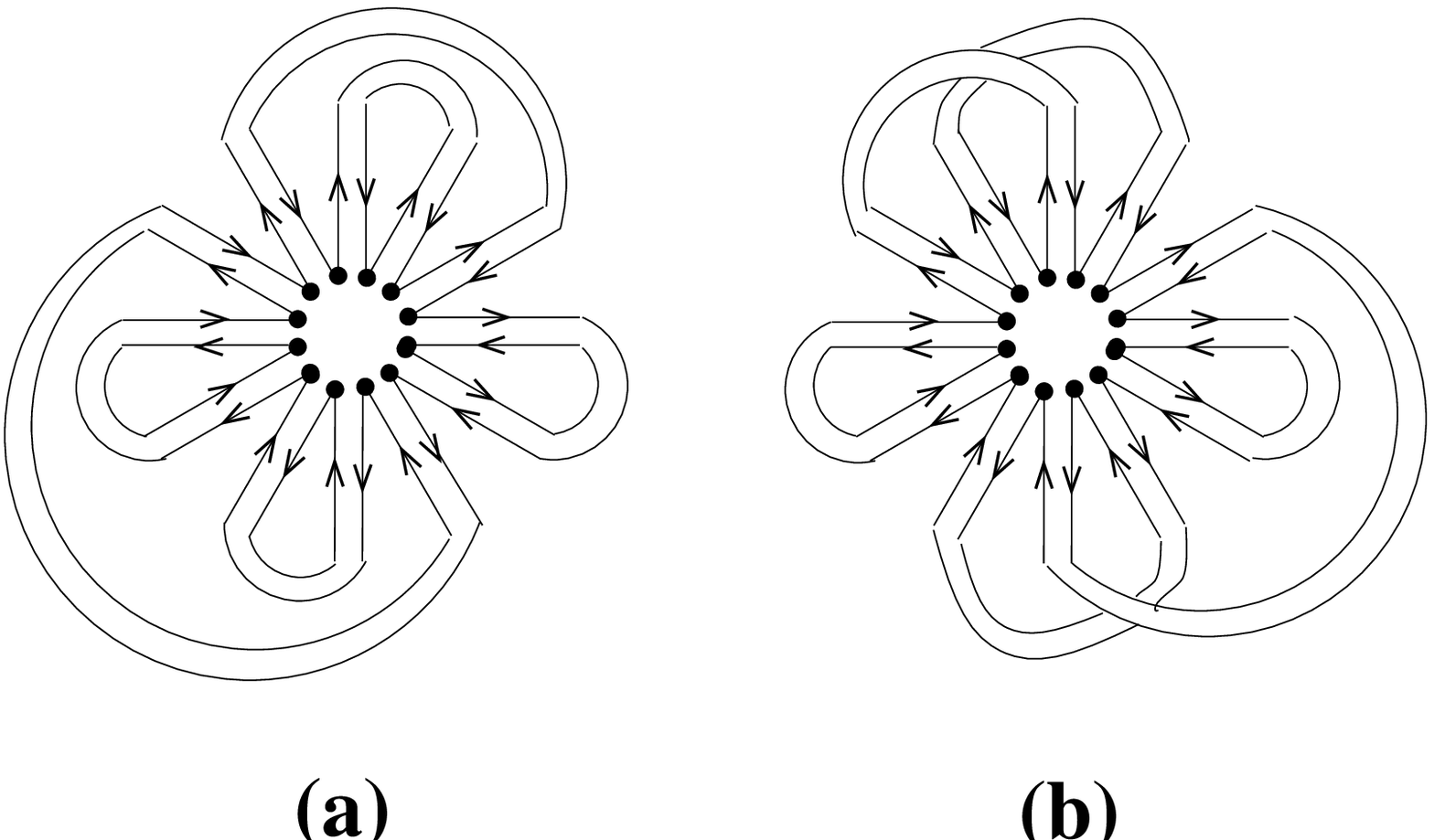}{7.cm}
\figlabel\planar

As a first application of this diagrammatic interpretation of the Wick
theorem \wimat, let us compute the large $N$ asymptotics of $\langle {\rm
Tr}(M^n)\rangle$. To compute $\langle {\rm
Tr}(M^n)\rangle$, we must first draw a star-diagram as in \tracrep, then
apply \wimat\ to express the result as a sum over the saturations of
the star with edges connecting its outcoming half-edges by pairs. 
To get a non-zero result, we must clearly have $n$ even, say $n=2p$.
Again,
there are $(2p-1)!!$ such pairings, and indeed we recover the case of 
previous section by taking $N=1$. But if instead we take $N$ to be 
large, we see that only a fraction of these $(2p-1)!!$ pairings will contribute
at leading order. Indeed, assume first we restrict the set of pairings to
{\it planar} ones (see Fig.\planar\ (a)), namely such that the saturated star diagrams
have a petal structure in which
the petals are  either juxtaposed or included into one-another (with no
edges-crossings). 
We may compute the genus of the petal diagrams by noting that they form
a tessellation of the sphere (=plane plus point at infinity). This tessellation
has $V=1$ vertex (the star), $E=p$ edges, and $F$ faces, including the
``external" face containing the point at infinity. The planarity of the diagram
simply expresses that its genus $h$ vanishes, namely
\eqn\plapro{ 2-2h=2=F-E+V=F+1-p \ \ \Rightarrow \ \ F=p+1}
Such diagrams
receive a total contribution $1/N^{p}$ from the propagators (weight $1/N$ 
per connecting edge), but we still have to sum over the remaining matrix
indices $j_1,j_2,...,j_{p+1}$ running over the $p+1$ oriented loops we have
created, which form the boundaries of the
$F=p+1$ faces.  This gives a weight $N$ per
face of the diagram, hence a total contribution of $N^{p+1}$. 
So all the petal diagrams
contribute the same total factor $N^{p+1}/N^p=N$ to $\langle {\rm Tr}(M^n)\rangle$. Now
any non-petal (i.e. non-planar, see Fig.\planar\ (b)) 
diagram must have at least {\it two less}
oriented loops. Indeed, its Euler characteristic is negative or zero, hence
it has $F\leq E-V=p-1$ and it contributes at most for $N^{F-p}\leq 1/N$.
So, to leading order in $N$, only the genus zero (petal) diagrams contribute. 
We simply have to count them. This is a standard problem in combinatorics:
one may for instance derive a recursion relation for the number $c_p$ of petal
diagrams with $2p$ half-edges, by fixing the left end of an edge (say at
position $1$), and summing over the positions of its right end (at positions
$2j$, $j=1,2,...,p$), and noting that the petal thus formed may contain
$c_{j-1}$ distinct petal diagrams and be next to $c_{p-j}$ distinct ones.
This gives the recursion relation
\eqn\recucat{ c_p=\sum_{j=1}^p c_{j-1} c_{p-j} \qquad c_0=1}
solved by the Catalan numbers
\eqn\catalan{ c_p={(2p)! \over (p+1)! p!} }
Finally, we get the one-matrix planar Gaussian average by taking the large $N$
limit:
\eqn\magau{ \lim_{N\to \infty} {1\over N} \langle {\rm Tr}(M^n) \rangle =
\left\{ \matrix{ c_p & {\rm if} \ \ n=2p \cr
0 & {\rm otherwise} \cr} \right. }
This exercise shows us what we have gained by considering $N\times N$ matrices
rather than numbers: we have now a way of discriminating between the various
genera of the graphs contributing to Gaussian averages. This fact will be
fully  exploited in the next example.

\subsec{Model building I: using one-matrix integrals}

Let us apply the matrix Wick theorem \wimat\ to the following generating
function $f(M)=\exp(N\sum_{i\geq 1} g_i {\rm Tr}(M^i)/i)$, to be understood
as a formal power series of the $g_i$, $i=1,2,3,4,...$
\eqn\gram{\eqalign{ Z_N(g_1,g_2,...)&=\langle e^{N\sum_{i\geq 1} 
g_i {\rm Tr}({M^i\over i})} \rangle\cr
&=\sum_{n_1,n_2,...\geq 0} \prod_{i\geq 1} {(Ng_i)^{n_i}\over i^{n_i} n_i!} \langle
\prod_{i\geq 1} {\rm Tr}(M^i)^{n_i} \rangle\cr
&=\sum_{n_1,n_2,...\geq 0}\prod_{i\geq 1} {(Ng_i)^{n_i}\over i^{n_i} n_i!} 
\sum_{{\rm all}\ {\rm labelled}\ {\rm fatgraphs}\ \Gamma \atop
{\rm with} \ n_i\ i-{\rm valent}\ {\rm vertices}} N^{-E(\Gamma)} N^{F(\Gamma)}\cr} }
by direct application \wimat. 
\fig{A typical connected fatgraph $\Gamma$, corresponding to the 
average $\langle {\rm Tr}(M)^3
{\rm Tr}(M^2)^2{\rm Tr}(M^3)
{\rm Tr}(M^4)^2 {\rm Tr}(M^6){\rm Tr}(M^8)\rangle$. The graph was obtained by 
saturating the ten star-diagrams corresponding to the ten trace terms,
namely with $n_1=3$ univalent vertices, $n_2=2$ bi-valent ones,
$n_3=1$ tri-valent one, $n_4=2$ four-valent ones, $n_6=1$ 
six-valent one and $n_8=1$ eight-valent one, hence a total of $V=10$ vertices. 
This graph corresponds to some
particular Wick pairing for which we have drawn the $E=16$ connecting edges,
giving rise to $F=2$ oriented loops bordering the faces of $\Gamma$.}{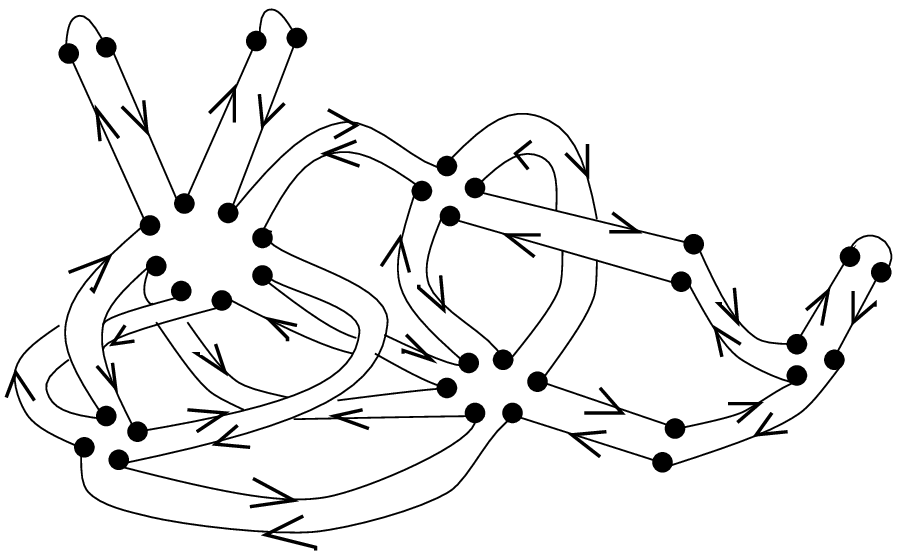}{8.cm}
\figlabel\gravi
\noindent In \gram,
we have first represented pictorially the integrand $\prod_i ({\rm
Tr}(M^i))^{n_i}$  as a succession of $n_i$ $i$-valent star diagrams like that
of \tracrep, $i=1,2,...$. Then we have summed 
over all possible saturations of all the marked half-edges of all these stars, 
thus forming (non-necessarily connected) ribbon or fatgraphs
$\Gamma$ with some labelling of their half-edges (see Fig.\gravi\ for an example 
of connected fatgraph). 
In \gram, we have  denoted by
$E(\Gamma)$ the total number of edges of $\Gamma$, connecting
half-edges by pairs, i.e. the number of propagators needed (yielding a factor
$1/N$ each, from \prop). The number $F(\Gamma)$ is the total number of faces of 
$\Gamma$. The faces of $\Gamma$ are indeed well-defined because $\Gamma$ is
a fatgraph, i.e. with edges made of doubly oriented parallel lines carrying the
corresponding matrix indices $i=1,2,...N$: the oriented loops we have
created by the pairing process are interpreted as face boundaries, in
one-to-one correspondence with faces of $\Gamma$. But the traces of the various
powers of $M$ still have to be taken, which means all the indices running from
$1$ to $N$ have to be summed over all these loops. This results in the factor
$N$ per face of $\Gamma$ in \gram. Finally, the  sum extends over all (possibly
disconnected) fatgraphs $\Gamma$ with labelled half-edges. Each such labelled
graph corresponds to exactly one Wick  pairing of \wimat. Summing over all the
possible labellings of a given un-labelled fatgraph $\Gamma$ results in some
partial cancellation of the symmetry prefactors $\prod_i 1/(i^{n_i}n_i!)$,
which actually leaves us with the inverse of the order of the symmetry group
of the un-labelled fatgraph $\Gamma$, denoted by $1/|Aut(\Gamma)|$. This gives
the final form
\eqn\fatag{ Z_N(g_1,g_2,...)=\sum_{{\rm fatgraphs}\atop \Gamma } \ 
{N^{V(\Gamma)-E(\Gamma)+F(\Gamma)}
\over |Aut(\Gamma)|}  
\prod_{i\geq 1} g_i^{n_i(\Gamma)} }
where $n_i(\Gamma)$ denotes the total number of $i$-valent vertices of $\Gamma$
and $V(\Gamma)=\sum_i n_i(\Gamma)$ is the total number of vertices of $\Gamma$.
To restrict the sum in \fatag\ to only connected graphs, we simply have to
formally expand the logarithm of $Z_N$, resulting in the final identity
\eqn\parfure{ F_N(g_1,g_2,...)={\rm Log}\, Z_N(g_1,g_2,...)=
\sum_{{\rm connected}\atop {\rm fatgraphs}\ \Gamma} 
{N^{2-2h(\Gamma)} \over |Aut(\Gamma)|}
\prod_i g_i^{n_i(\Gamma)} }
where we have identified the Euler characteristic $\chi(\Gamma)=F-E+V=2-2h(\Gamma)$,
where $h(\Gamma)$ is the genus of $\Gamma$ (number of handles).
Eqn.\parfure\ gives a clear geometrical meaning to the Gaussian average of our
choice of $f(M)$: it amounts to computing the generating function for fatgraphs
of given genus and given vertex valencies. 
Such a fatgraph $\Gamma$ is in turn dual to a 
tessellation $\Gamma^*$ of a Riemann surface of same genus,
by means of $n_i$ $i$-valent polygonal tiles, $i=1,2,...$. 

The result \parfure\ is therefore a statistical sum over discretized random
surfaces (the tessellations), that can be interpreted in physical terms as the
free energy of a model of discrete 2D quantum gravity. It simply identifies the Gaussian
matrix integral with integrand $f(M)$ as a discrete sum over configurations of
tessellated surfaces of arbitrary genera, weighted by some exponential 
factor. More precisely, imagine only $g_3=g\neq 0$ while all other $g_i$'s vanish.
Then \parfure\ becomes a sum over fatgraphs with cubic (or 3-valent) vertices, dual 
to triangulations $T$ of Riemann surfaces of arbitrary
genera. Assuming these triangles have all unit area, then $n_3(\Gamma)=A(T)$ is
simply the total area of the triangulation $T$. Hence \parfure\ becomes
\eqn\furpar{ F_N(g)=\sum_{{\rm connected}\ {\rm triangulations}\ T} 
{g^{A(T)} N^{2-2h(T)}\over |Aut(T)|} }
and the summand $g^A N^{2-2h}=e^{-S_E}$ is nothing but the exponential of the 
discrete version of Einstein's
action for General Relativity in 2 dimensions \generel,
in which we have identified the two invariants of $\Sigma$: its area $A(\Sigma)$ and its
Euler characteristic $\chi(\Sigma)=2-2h(\Sigma)$. 
The contact
with \furpar\ is made by setting $g=e^{-\Lambda}$ and $N=e^{-{\cal N}}$.

\fig{A 4-valent planar graph with hard dimers, represented by thickened
edges. The corresponding graph obtained by shrinking the dimers (b) has both 4-valent
and 6-valent vertices. The correspondence is three-to-one per dimer, as shown.}{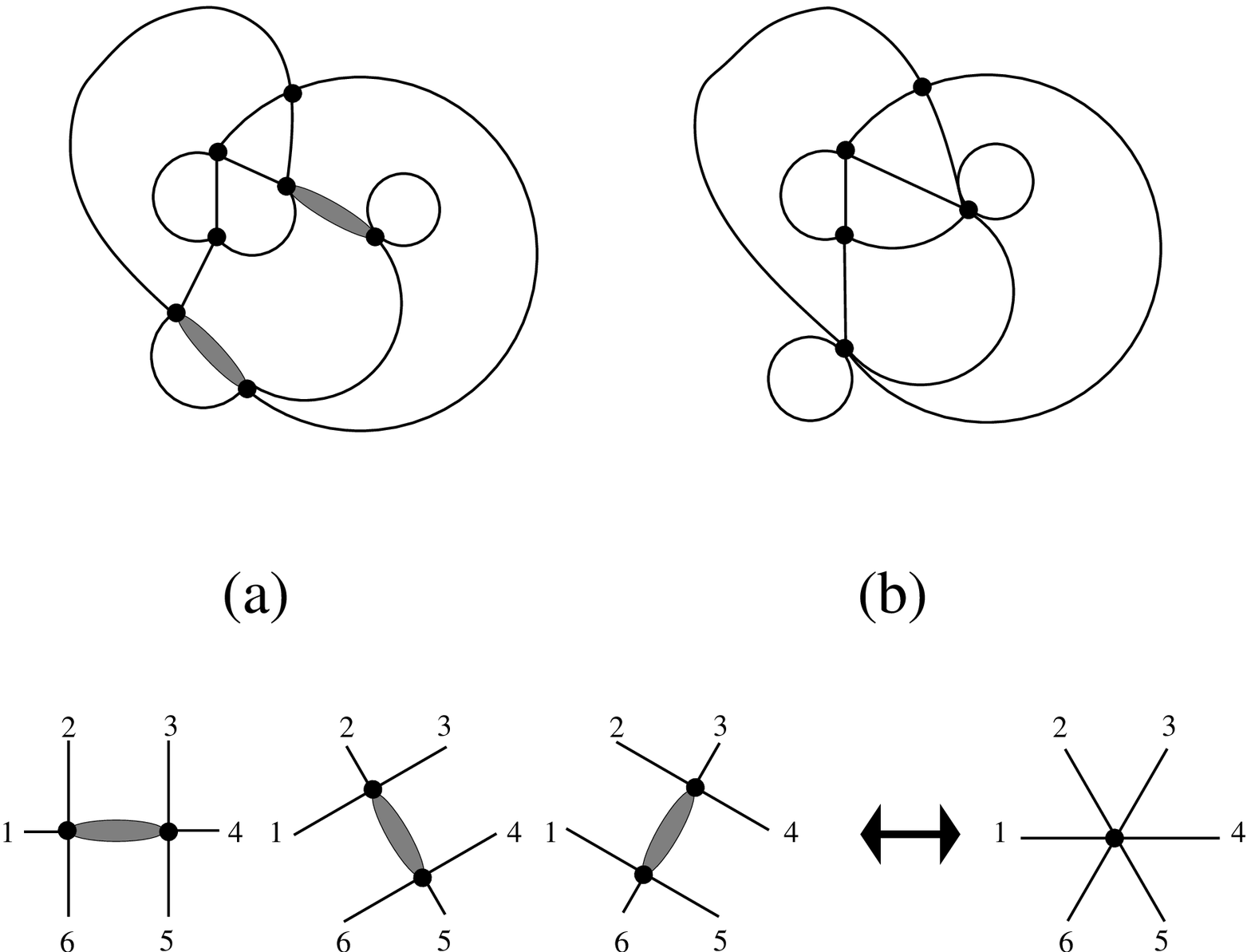}{10.cm}
\figlabel\dimer

If we now include all $g_i$'s in \parfure\ we simply get a more elaborate
discretized model, in which we can keep track of the valencies of vertices of
$\Gamma$ (or tiles of the dual $\Gamma^*$). These in turn may be understood
as discrete models of matter coupled to 2D quantum gravity.
This is best seen in the case of the Hard-Dimer model 
on random 4-valent graphs \STAU. The configurations
of the model are made of arbitrary 4-valent fatgraphs of arbitrary genus
(the underlying discrete fluctuating space) and of choices of edges
occupied by dimers, with the hard-core condition that no two adjacent edges
may be simultaneously occupied (see Fig.\dimer\ for an illustration in the case of 
a planar graph). 
These matter configurations are given an
occupation energy weight $z$ per dimer, while the space part receives
the standard weight $g$ per 4-valent vertex, and the overall
weight $N^{2-2h}$ for each graph of genus $h$. We then note that any occupied dimer
may be shrunk to naught, thus creating a 6-valent vertex by the fusion of
its two 4-valent adjacent vertices. Comparing the configurations of the Hard-Dimer model
on 4-valent graphs and those of graphs with only 4-  and 6-valent vertices, we see that there
is a one-to-three correspondence between those, as there are exactly three ways of
decomposing a 6-valent vertex into two adjacent 4-valent ones connected by a dimer
(see the bottom line of Fig.\dimer).
The Hard-Dimer model is therefore generated by an integral of the form \fatag, with
only $g_4$ and $g_6$ non-zero, and more precisely $g_4=g$ and $g_6=3 g^2 z$
(=three decompositions into two 4-valent vertices and one dimer). This is the simplest 
instance of matter coupled to 2D quantum gravity we could think of, and it indeed corresponds 
to graphs with specific valence weights.

Going back to the purely mathematical interpretation of \parfure, we start to
feel how simple matrix integrals can be used as tools for generating 
all sorts of graphs whose duals tessellate surfaces of arbitrary given 
topology. The size $N$ of the matrix relates to the genus, whereas the
details of the integrand relate to the structure of vertices. An important 
remark is also that the large $N$ limit of \parfure\ extracts the genus zero
contribution, namely that of planar graphs. So as a by-product,
it will be possible to extract results on planar graphs from asymptotics of 
matrix integrals for large size $N$.

\subsec{Model building II: using multi-matrix integrals}

The results of previous section can be easily generalized to multiple Gaussian
integrals over several Hermitian matrices. More precisely, let $M_1$, $M_2$,
... $M_p$ denote $p$ Hermitian matrices of same size $N\times N$, and
$Q_{a,b}$, $a,b=1,2,...,p$ the elements of a positive definite form $Q$.
We consider the multiple Gaussian integrals of the form
\eqn\fromult{ \langle f(M_1,...,M_p)\rangle ={\int dM_1...dM_p e^{-{N\over 2}
\sum_{a,b=1}^p {\rm Tr}(M_a Q_{ab}M_b) } f(M_1,...,M_p) \over 
\int dM_1...dM_p e^{-{N\over 2}
\sum_{a,b=1}^p {\rm Tr}(M_a Q_{ab}M_b) } } }
The one-Hermitian matrix case of the previous section
corresponds simply to $p=1$ and $Q_{1,1}=1$. The averages \fromult\
are computed by extending the
source integral method of previous section: for some Hermitian source matrices
$S_1,...,S_p$ of size $N\times N$, we define and compute the
multi-source integral 
\eqn\soucmu{\Sigma(S_1,...,S_p)= 
\langle e^{\sum_{a=1}^p {\rm Tr}(S_a M_a)} \rangle =
e^{{1 \over 2N} \sum_{a,b=1}^p {\rm Tr}(S_a (Q^{-1})_{a,b} S_b)} }
and apply multiple derivatives w.r.t. to $S_a$'s to compute any expression of the form
\fromult, before taking $S_a\to 0$. As before, derivatives w.r.t. elements of the
$S$'s must go by pairs to yield a non-zero result. For instance, in the case
of two matrix elements of $M_a$'s we find the propagators
\eqn\propmu{ \langle (M_a)_{ij} (M_b)_{kl}\rangle ={1\over N} \delta_{il} \delta_{jk} 
(Q^{-1})_{a,b} }
In general we will apply the multi-matrix Wick theorem
\eqn\muwic{ \langle \prod_{(a,i,j)\in J} (M_a)_{ij} \rangle=
\sum_{{\rm pairings}\atop P} \prod_{{\rm pairs}\atop
(aij),(bkl)\in P} \langle (M_a)_{ij} (M_b)_{kl}\rangle }
expressing the multi-matrix Gaussian average of any product of matrix elements of the
$M$'s as a sum over all pairings saturating the matrix half-edges, weighted 
by the corresponding value of the propagator \propmu. Note that half-edges 
must still be connected according to the rule \propag, but that in addition,
depending on the form of $Q$, some matrices may not be allowed to connect 
to one another (e.g. if $(Q^{-1})_{ab}=0$ for some $a$ and $b$,
then $\langle M_a M_b\rangle=0$,
and in such a case, there cannot be any edge connecting a matrix with index $a$ to one
with index $b$).

This gives us much freedom in cooking up multi-matrix models to evaluate generating 
functions of graphs with specific decorations such as colorings, spin models,
etc... This is expected to describe the coupling of matter systems (e.g.
a spin model usually defined on a regular lattice) to 2D quantum gravity (by
letting the lattice fluctuate into tessellations of arbitrary genera).
Famous examples are the O(n) model \ON, the q-states Potts model \QPO, 
both including the Ising model as particular cases. Other models of interest require
to use different types of matrices, to best represent their degrees of freedom.
This is the case for the 6 vertex model expressed in terms of complex matrices, and 
for the so-called IRF (interaction round a face) models, expressed in terms of 
complex rectangular arrays \IRFK\ \RECT.

\newsec{The one-matrix model I: large $N$ limit and the enumeration of planar graphs}

In this section, we will
mainly cover the one-matrix integrals defined in Sect.2.4. Multi-matrix
techniques are very similar, and we will present them in a concluding section.
More precisely, we will study the one-matrix integral
\eqn\onmav{ Z_N(V)={\int dM e^{-N{\rm Tr}\, V(M)} \over
\int dM e^{-N{\rm Tr}\, V_0(M)} }}
with an arbitrary polynomial potential, say
\eqn\arbipo{V(x)={x^2\over 2}- \sum_{i=1}^d {g_i\over i} x^i,\quad {\rm and}\quad
V_0(x)={x^2\over 2}}
This contains as a limiting case the partition function \gram\ of Sect.2.4.
Note also that we are not worrying at this point about convergence issues for these
integrals, as they must be understood as formal tools allowing for computing
well-defined coefficients in formal series expansions in the $g$'s. 

\subsec{Eigenvalue reduction}

The step zero in computing the integral \onmav\ is the reduction to a $N$-dimensional 
integral, namely over the real eigenvalues $m_1,...,m_N$ of the Hermitian matrix $M$.
This is done by performing the change of variables $M\to (m,U)$, where 
$m=$diag$(m_1,...,m_N)$, and $U$ is a unitary diagonalization matrix such that
$M=UmU^\dagger$, hence $U\in U(N)/U(1)^N$ as $U$ may be multiplied by an arbitrary
matrix of phases. The Jacobian of the transformation is readily found to be
the squared Vandermonde determinant
\eqn\jaco{ J= \Delta(m)^2 = \prod_{1\leq i<j\leq N} (m_i-m_j)^2 }
A simple derivation consists in expressing the differential $dM$ in terms
of $dU$ and $dm$ in the vicinity of $U=I$, namely $dM=dU m+dm+mdU^\dagger$, but
noting that $UU^\dagger =I$, we get $dU^\dagger=-dU$, and finally
$dM=dm+[dU,m]$, or $dM_{ij}=dm_i\delta_{ij}+(m_i-m_j)dU_{ij}$, from which we directly
read the Jacobian \jaco.
Performing the change of variables in both the numerator and denominator of \onmav\ 
we obtain
\eqn\obtz{ Z_N(V)={\int_{\IR^N} dm_1...dm_N \Delta(m)^2 e^{-N\sum_{i=1}^N V(m_i)}
\over \int_{\IR^N} dm_1...dm_N \Delta(m)^2 e^{-N\sum_{i=1}^N {m_i^2\over 2}} } }

\subsec{Large size: the saddle-point technique}

Starting from the 
$N$-dimensional integral \obtz, we rewrite
\eqn\robtz{ Z_N(V)={\int dm_1...dm_N e^{-N^2 S(m_1,...,m_N)} \over
\int dm_1...dm_N e^{-N^2 S_0(m_1,...,m_N)}} }
where we have introduced the actions
\eqn\actio{ \eqalign{
S(m_1,...,m_N)&= {1 \over N}\sum_{i=1}^NV(m_i) -{1\over N^2}\sum_{1\leq i\neq j\leq N}
{\rm Log}\vert m_i-m_j\vert \cr
S_0(m_1,...,m_N)&={1 \over N}\sum_{i=1}^NV_0(m_i) -{1\over N^2}\sum_{1\leq i\neq j\leq N}
{\rm Log}\vert m_i-m_j\vert \cr}}
For large $N$ the numerator and denominator of \robtz\ are dominated by the 
semi-classical (or saddle-point) minima of $S$ and $S_0$ respectively.
For $S$, the saddle-point equations read
\eqn\sap{ {\partial S\over\partial m_j}=0 \ \Rightarrow \ 
V'(m_j)= {2 \over N} \sum_{1\leq i \leq N\atop i\neq j} {1 \over m_j-m_i} }
for $j=1,2,...,N$.
Introducing the discrete resolvent 
\eqn\dires{ \omega_N(z)={1\over N} \sum_{i=1}^N {1\over z-m_i} }
evaluated at the solution $m_1,..,m_N$ to \sap,
multiplying \sap\ by $1/(N(z-m_j))$ and summing over $j$, 
we easily get the equation
\eqn\qares{\eqalign{
V'(z) \omega_N(z)&+{1\over N}\sum_{j=1}^N {V'(m_j)-V'(z)\over z-m_j}\cr
&={1\over N^2} \sum_{1\leq i\neq j\leq N} 
{1\over m_j-m_i}\bigg({1\over z-m_j}-{1\over z-m_i}
\bigg) \cr
&={1\over N^2} \sum_{1\leq i\neq j\leq N} {1\over (z-m_i)(z-m_j)}\cr
&=\omega_N(z)^2+{1\over N}\omega_N'(z) \cr}}
Assuming $\omega_N$ tends to a differentiable function $\omega(z)$ when $N\to \infty$ 
we may neglect the last derivative term, and we are left with the quadratic equation
\eqn\qadrares{\eqalign{ &\omega(z)^2 -V'(z) \omega(z)+ P(z) =0 \cr
&P(z)=\lim_{N\to \infty}{1\over N}\sum_{j=1}^N {V'(z)-V'(m_j)\over z-m_j} \cr}}
where $P(z)$ is a polynomial of degree $d-2$, $d$ the degree of $V$. The existence of the limiting resolvent
$\omega(z)$ boils down to that of the limiting density of distribution of
eigenvalues
\eqn\lidens{ \rho(z)=\lim_{N\to \infty} {1\over N}\sum_{j=1}^N \delta(z-m_j)}
normalized by the condition 
\eqn\normaro{ \int_\IR \rho(z) dz =1 }
as there are exactly $N$ eigenvalues on the real axis. This density is related 
to the resolvent through
\eqn\resden{ \omega(z)=\int {\rho(x)\over z-x} dx=
\sum_{m=1}^\infty {1\over z^m} \int_\IR x^{m-1}\rho(x) dx } 
where the expansion holds in the large $z$ limit, and the integral extends over 
the support of $\rho$, included in the real line. 
Conversely, the density is obtained from the resolvent by use of the 
discontinuity equation across its real support 
\eqn\discon{ \rho(z)={1\over 2 i \pi}\lim_{\epsilon\to 0} \omega(z+i\epsilon)-
\omega(z-i\epsilon) \qquad z\in {\rm supp}(\rho) }
Solving the quadratic equation \qadrares\ as
\eqn\solqa{ \omega(z)= {V'(z)-\sqrt{ (V'(z))^2 -4 P(z)} \over 2} }
we must impose the large $z$ behavior inherited from \normaro\-\resden, namely
that $\omega(z)\sim 1/z$ for large $z$. For $d\geq 2$, the polynomial in the square 
root has degree $2(d-1)$: expanding the square root for large $z$ up to order $1/z$,
all the terms cancel up to order $0$ with $V'(z)$,
and moreover the coefficient in front of $1/z$ must be $1$ (this fixes the leading 
coefficient of $P$). The other coefficients of $P$ are fixed by the higher moments of
the measure $\rho(x) dx$. 

For instance, when $k=2$ and $V=V_0$, we get $P=1$ and
\eqn\rezer{ \omega_0(z)= {1\over 2} (z-\sqrt{z^2-4} ) }
It then follows from \discon\ that the density has the compact support $[-2,2]$ and has the
celebrated ``Wigner's semi-circle law" form
\eqn\forro{ \rho_0(z)= {1\over 2\pi } \sqrt{4-z^2} } 
The resolvent $\omega_0$ is the generating function for the moments of the 
measure whose density is $\rho_0$ (via the expansion \resden), from which we 
immediately identify 
\eqn\mom{ \int_\IR x^n \rho_0(x) dx = \left\{ \matrix{ c_p & {\rm if}\ n=2p \cr
0 & {\rm otherwise} \cr} \right. }
with $c_p$ as in \catalan. Indeed, due to the quadratic recursion relation
\recucat, the generating function $C(x)=\sum_{p\geq 0} x^p c_p$ satisfies $xC(x)^2=C(x)-1$,
and therefore we have $\omega_0(z)=C(1/z^2)/z$.
The coefficients \mom\ are nothing but the planar limit of the Gaussian Hermitian matrix averages 
(with potential $V_0(x)=x^2/2$), namely 
$\lim_{N\to \infty}\langle {1\over N}{\rm Tr}M^n\rangle_{V_0}
=\int_\IR x^n \rho_0(x) dx$,
hence our analytical result \mom\ is an alternative for that already obtained 
combinatorially in \magau. 

In the general case, the density reads 
\eqn\geneden{ \rho(z)= {1\over 2\pi}\sqrt{4P(z)-(V'(z))^2} }
and may have a disconnected support, made of a union of intervals 
(the so-called multicut solutions). 
It is however interesting to restrict oneself to the case when the support of $\rho$ is
made of a single real interval $[a,b]$, as this will always be the preferred
saddle-point solution for generating the correct formal series expansions of the
all-genus free energy. For supports made of more than one interval, resonances may occur
as eigenvalues tunnel from one interval to another, and oscillations develop in the $N$
dependence, which cause the large $N$ expansion to break down, unless some strong conditions are
imposed on say complex contour integrals for the eigenvalues. 
The one-cut hypothesis will
be justified {\it a posteriori} in Sect.4 below, when we revisit the problem from
a purely combinatorial perspective.

In the one-cut case, the polynomial $V'(z)^2-4P(z)$ has 
single roots at say $z=a$ and $z=b$ and all other roots have even multiplicities.
In other words, we may write the limiting resolvent as
\eqn\omez{\omega(z)={1\over 2}(V'(z) -Q(z)\sqrt{(z-a)(z-b)} )}
where $Q(z)$ is a polynomial of degree $k-2$, entirely fixed in terms of $V$ by the 
asymptotics $\omega(z)\sim 1/z$ for large $|z|$. More precisely, let us introduce
$H(z)=V'(z)/\sqrt{(z-a)(z-b)}$, considered as a series expansion for large $z$, then
$Q(z)$ is nothing but the part of this series that is polynomial in $z$,
denoted as $H_+(z)$. Writing moreover $H(z)=H_+(z)+H_-(z)$, we finally get
\eqn\omval{ \omega(z)={1\over 2}H_-(z)\sqrt{(z-a)(z-b)} }
Writing $H_-(z)=\sum_{i\geq 1} H_{-i} z^{-i}$, we get that $\omega(z)\sim 1/z$ iff
$H_{-1}=0$ and $H_{-2}=2$. These coefficients are expressed as residue integrals
at infinity, namely
\eqn\resintinf{ H_{-m}(z)=\oint {dz \over 2i\pi}z^{m-1} { V'(z) \over \sqrt{(z-a)(z-b)}} }
The square root term is uniformized by the change of variables
$z=w+S+R/w$, with $S={a+b\over 2}$ and $R=\left({b-a\over 4}\right)^2$, and
\eqn\eqinthm{ H_{-m}(z)=\oint {dw\over 2i\pi w} (w+S+R/w)^{m-1} V'(w+S+R/w) }
so that finally $H_{-1}=V'_{0}$ and $H_{-2}=V'_{-1}+S 
V'_{0}+RV'_{1}$, where the shorthand notation $V'_m$ stands for the coefficient of
$w^m$ in the large $w$ expansion of $V'(w+S+r/w)$.
Performing the change of variables $w\to R/w$ allows to relate 
$V'_{-m}=R^mV'_{m}$. Finally, the asymptotic
condition $\omega(z)=1/z+O(1/z^2)$ at large $z$ boils down to
\eqn\rssystem{\eqalign{ V'_{0}&=0=S-\sum_{i\geq 1}g_i \sum_{j=0}^{[(i-1)/2]} S^{i-2j-1} 
R^j{(i-1)!\over (j!)^2(i-2j-1)!}\cr
V'_{-1}&=1=R-\sum_{i\geq 1}g_i \sum_{j=0}^{[i/2]}S^{i-2j} 
R^j {(i-1)!\over j!(j-1)!(i-2j)!}\cr}}
These equations simplify drastically in the case of even potentials, where $g_i=0$ for all
odd $i$. The parity of $V$ indeed induces that of $\rho$, and we have $S=(a+b)/2=0$ as the support
of the density is symmetric w.r.t. the origin. This leaves us with only one equation
\eqn\onlyone{ 1=R-\sum_{i\geq 1} g_{2i}R^i {2i-1\choose i} }
for $R=a^2/4$.
In the particular case of the gaussian potential $V=V_0$, this reduces to $R=1$ and $S=0$,
in agreement with $b=-a=2$ \rezer. Expanding the solutions of \rssystem\ as formal power series
of the $g_i$'s, the conditions $R=1+O(\{g_i\})$ and $S=O(\{g_i\})$ determine them uniquely. These
in turn determine $a$ and $b$ and therefore $\rho$ and $\omega$ completely.

The planar free energy $f=F-F_0=\lim_{N\to \infty} {1\over N^2} {\rm Log}\, \big(Z_N(V)/Z_N(V_0)\big)$
is finally obtained by substituting the limiting densities $\rho,\rho_0$ in the 
saddle point actions $S$ and $S_0$, with the result $F-F_0=S_0-S$.
It is however much simpler to evaluate some derivatives of the free energy, by directly
relating them to the planar resolvent $\omega(z)$, the subject of next section.

\subsec{Enumeration of planar graphs with external legs}

\fig{Samples of planar graphs with external legs (univalent vertices
marked with a cross) and arbitrary valences, with respectively 
(a) one leg in the external face (b) one leg (anywhere)
(c) two legs in the same (external) face (d) two-legs (one in the external face, 
the other anywhere).}{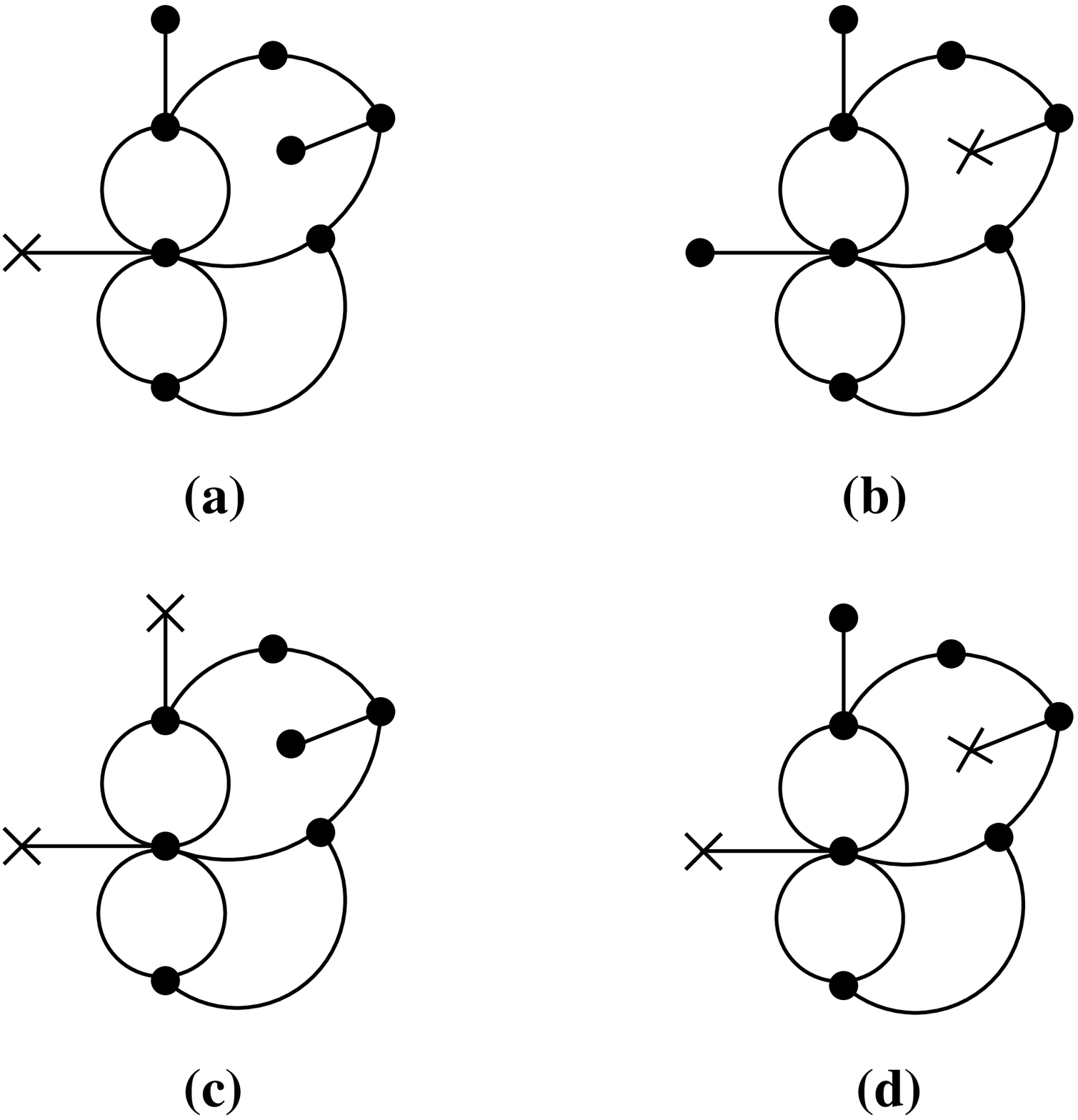}{10.cm}
\figlabel\planargraphs

Let us first consider the
generating function $\Gamma_1$
for planar graphs with weights $g_i$ per i-valent vertex, and with one
external (univalent) leg, represented in the external face on the plane (see Fig.\planargraphs\ (a)):
\eqn\comderiv{\Gamma_1=\partial f/\partial g_1=\lim_{N\to\infty}{1\over N}\langle
{\rm Tr}(M)\rangle_V=\omega_{-2} ={1\over 2}(H_{-3}-SH_{-2}-2RH_{-1}) }
where use has been made of \resden,
and as before $\omega_{-m}$ denotes the coefficient of $z^{-m}$ in the large 
$z$ expansion of $\omega(z)$. From the large $z$ asymptotics of $\omega(z)$, we know that
$H_{-1}=0$ and $H_{-2}=2$, and we must now evaluate $H_{-3}=V'_{-2}+2SV'_{-1}+(S^2+2R)V'_{0}+2RSV'_{1}+
R^2V'_2=2(V'_{-2}+2SV'_{-1})+(S^2+2R)V'_{0}=2(V'_{-2}+2S)$, leaving us with
\eqn\gammaone{\Gamma_1=V'_{-2}+S}
Analogously, we may compute the connected two-leg-in-the-same-face graph generating 
function $\Gamma_2=\omega_{-3}-\Gamma_1^2$ (see Fig.\planargraphs\ (c)),
in which we subtract the contributions from disconnected pairs of one-leg graphs. We get
$\omega_{-3}=\partial f/\partial g_2=R+S^2+V'_{-3}+2S V'_{-2}$ and finally
\eqn\gammatwo{\Gamma_2=R+V'_{-3}-(V'_{-2})^2 }

Another quantity of interest is the connected two-leg graph generating 
function $\Gamma_{1,1}=\partial^2 f/\partial g_1^2=\partial \omega_{-2}/\partial g_1$
(see Fig.\planargraphs\ (d)).
This turns into
\eqn\calcgaoneone{
\Gamma_{1,1}={ \partial\omega_{-2}\over \partial g_1}={ \partial S\over \partial g_1}+
{ \partial V'_{-2}\over \partial g_1}
={ \partial S\over \partial g_1}(1+V''_{-2}) +{\partial R\over \partial g_1} V''_{-1}}
Let us first replace the term $1$ in factor of $\partial S/\partial g_1$
by $1=V'_{-1}$, the second equation of \rssystem.
Note that the residue of a total differential always vanishes, hence in particular
\eqn\trick{\big( d/dw(w V'(w+S+R/w))\big)_{-1}=0=V'_{-1}+V''_{-2}-RV''_{0}}
This allows to rewrite
\eqn\rewoneone{
\Gamma_{1,1}={ \partial S\over \partial g_1} R V''_{0}+{\partial R\over \partial g_1} V''_{-1}}
Finally, differentiating the equation $V'_{0}=0$ w.r.t. $g_1$ yields 
$0=\partial S/\partial g_1 V''_{0}+\partial R/\partial g_1 V''_{1}-1$, where the last term 
comes from the explicit derivation w.r.t. $g_1$ of $V'(x)=x-g_1-g_2x-g_3x^2/2-...$
Multiplying this by $R$, and noting as before that $RV''_{1}=V''_{-1}$, we get
$R\partial S/\partial g_1 V''_{0}+\partial R/\partial g_1 V''_{-1}=R$ and finally
\eqn\lutfingaone{ \Gamma_{1,1}=R }
This result holds for even potentials as well, upon setting all $g_{2i+1}=0$ in the end.
Eq.\lutfingaone\ gives a straightforward combinatorial interpretation of $R$ as the generating function
for planar graphs with two external (univalent) legs, not necessarily in the same face.

To conclude the section, let us now give a combinatorial interpretation for $S$.
Let us show that $S$ is the generating function for one-leg planar graphs. By this we mean that the
leg need not be adjacent to the external face, as was the case for $\Gamma_1$ (see Fig.\planargraphs\ (b)). 
Comparing with the definition of $\Gamma_1$, we must show that $S$ is the generating function
for one-leg planar graphs (with the leg in the external face), and with a marked face
(chosen to be the new external face).
This amounts to the identity 
\eqn\idenSGa{S=z\partial_z \Gamma_1\vert_{z=1}}
where we have included a weight $z$ per {\it face} of the graph, to be set to $1$ in
the end. Due to Euler's relation
$F=2+E-V$, where $E$ is the total number of edges, and $V$ that of vertices of the one-leg
graphs at hand, and noting that $2E=1+\sum i V_i$ while $V=1+\sum V_i$, where $V_i$ is the number of
internal $i$-valent vertices, so that $2E-V=\sum (i-1)V_i$,
we see that $z\partial_z\Gamma_1=(2+t\partial_t)\Gamma_1$, if we attach a weight 
$1/t$ per edge and $t^{i-1}$ per $i$-valent vertex (with a net resulting weight $t^{2E-V-E}=t^{E-V}$).
Modifying the propagator and vertex weights of the matrix model accordingly,
this simply amounts to replacing $V'(x)$ by $V'(t x)=tx -\sum g_i t^{i-1} x^{i-1}$ 
in all the above formulas, and setting $t=1$ after differentiation.
This yields 
\eqn\wriST{ (2+\partial_t)\Gamma_1\vert_{t=1}=2 S+2 V'_{-2}
+{\partial S\over \partial t}\vert_{t=1}(1+V''_{-2})
+{\partial R\over \partial t}\vert_{t=1}V''_{-1}+V''_{-3}+SV''_{-2}+RV''_{-1} }
We now use the above trick \trick\ that the residue of a derivative vanishes, but this time
with
\eqn\newtrick{ \big( d/dw(w^2 V'(w+S+R/w))\big)_{-1}=0=2V'_{-2}+V''_{-3}-RV''_{-1}}
and we use this to eliminate $V''_{-3}$ from \wriST, as well as \trick\ to rewrite the
factor of $\partial S/\partial t$ as $1+V''_{-2}=V'_{-1}+V''_{-2}=RV''_0$, with the result
\eqn\resinterSG{ (2+\partial_t)\Gamma_1\vert_{t=1}=2 S+SV''_{-2}+2RV''_{-1}
+RV''_0{\partial S\over \partial t}\vert_{t=1}+V''_{-1}{\partial R\over \partial t}\vert_{t=1}}
Let us now differentiate w.r.t. $t$ the equation $0=V'_0$, and then set $t=1$ and multiply it by $R$. 
This gives
\eqn\intertrick{\eqalign{ 
0&=R(V''_{-1}+SV''_0+R V''_1+V''_0{\partial S\over \partial t}\vert_{t=1}+
V''_1 {\partial R\over \partial t}\vert_{t=1})\cr
&= RSV''_0+2RV''_{-1}+RV''_0{\partial S\over \partial t}\vert_{t=1}+
V''_{-1} {\partial R\over \partial t}\vert_{t=1}\cr}}
and allows to rewrite \resinterSG\ as
\eqn\presque{ (2+\partial_t)\Gamma_1\vert_{t=1}=2 S+SV''_{-2}-RSV''_0=S+S(V'_{-1}+V''_{-2}-RV''_0)=S}
by replacing $1\to V'_{-1}$ and using again the equation \trick.
This completes the identification of $S$ as the generating function for one-leg planar graphs,
with the leg not necessarily in the external face.

That the generating functions for both one- and two-leg planar graphs 
should satisfy a system of two algebraic equations \rssystem, looks like magic at first sight. 
It is the purpose of Sect.4 below to unearth the combinatorial grounds for this apparent miracle.

\subsec{The case of 4-valent planar graphs}

Before going into this, let us conclude with the case of the quartic potential 
say $V(z)={z^2\over 2}-g {z^4\over 4}$, for which we have $S=0$
and eq.\onlyone\ reduces to 
\eqn\quareduce{ 1=R-3gR^2 \quad \Rightarrow \quad R={a^2\over 4}={1\over 6g}(1-\sqrt{1-12 g}) }
as $R$ is the unique solution with the power series expansion $R=1+O(g)$. The corresponding resolvent
and density of eigenvalues read respectively
\eqn\evenquart{\eqalign{
\omega(z)&={1\over 2}(z-g z^3 -(1-g{a^2\over 2}-g z^2)\sqrt{z^2-a^2})\cr
\rho(z)&={1\over 2 \pi}(1-g{a^2\over 2}-g z^2)\sqrt{a^2-z^2} \cr}}
The two-leg-in-the-same-face graph generating function $\Gamma_2$ of eq.\gammatwo\ reads here
\eqn\quagamto{ \Gamma_2=R-gR^3={R(4-R)\over 3} }
where we have used eq.\quareduce\ to eliminate $g$. But any planar 4-valent graph with two external
legs in the same face is obtained by cutting an arbitrary edge in any closed planar 4-valent graph.
As the two legs are distinguished, and as there are exactly twice as many edges than vertices
in a closed 4-valent graph, we have $\Gamma_2=1+4g\partial f/\partial g$. The contribution $1$
comes from the unique graph made of one loop, with one edge and no vertex, not counted in $f$.
This gives the differential equation
\eqn\difequa{ 4 g {d f\over d g}={(R-1)(3-R)\over 3} }
and eliminating $g=(R-1)/(3R^2)$ from \quareduce, we finally get ${d f\over d g}=R^2(3-R)/4$.
Changing variables to $R$, this turns into ${d f\over d R}=(2-R)(3-R)/(12R)$, easily
integrated into
\eqn\easyintqua{ f={1\over 2} {\rm Log}\, R +{1\over 24} (R-1)(R-9) }
where the constant of integration is fixed by requiring that $f=0$ when $R=1$ (Gaussian case $V=V_0$).
Substituting the expansion $R=1+3g+18 g^2+...$ into \easyintqua\ yields the
expansion
\eqn\expanqua{f={g\over 2}+ \qquad {9\over 8} g^2\qquad +... }
$$\figbox{7.cm}{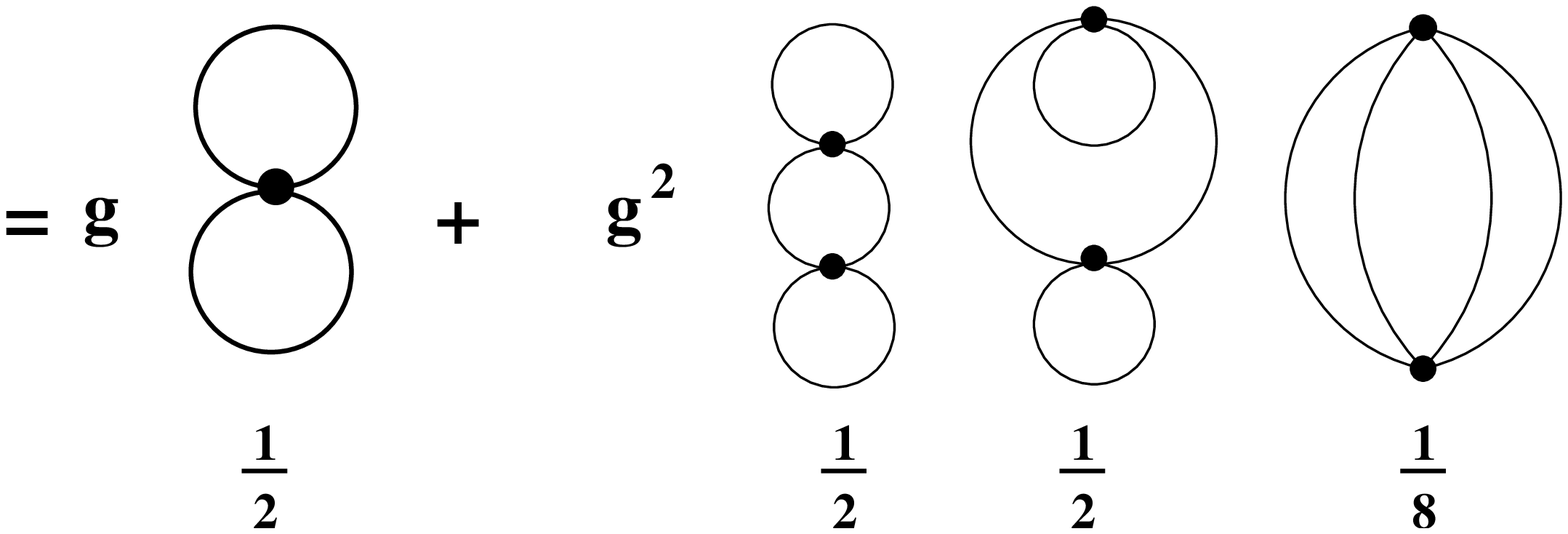}$$
where we have represented the planar 4-valent graphs with up to 2 vertices, together
with their inverse symmetry factors.

\newsec{The trees behind the graphs}

Using the above interpretation of $R$ as the generating function for planar graphs
with two distinguished external legs not necessarily in the same face, let us now
establish a general bijection between such graphs and suitably decorated trees,
also called blossom-trees. 

\subsec{4-valent planar graphs and blossom trees}

\fig{Illustration of the bijection between two-leg planar 4-valent graphs
and rooted blossom trees. Starting from a two-leg graph (a), we apply the iterative 
cutting procedure, which here requires turning twice around
the graph. In (b), the indices indicate the order in which the edges are cut during the 
$1$st turn ($1,2,3$) and $2$nd turn ($4,5,6$). Each cut edge is replaced by a black/white leaf pair
(c), while the in-coming leg is replaced by a leaf and the out-coming one by a root,
finally leading to a blossom tree (d). Conversely, the matching of black and white leaves
of the blossom tree (d) rebuilds the edges of (a).}{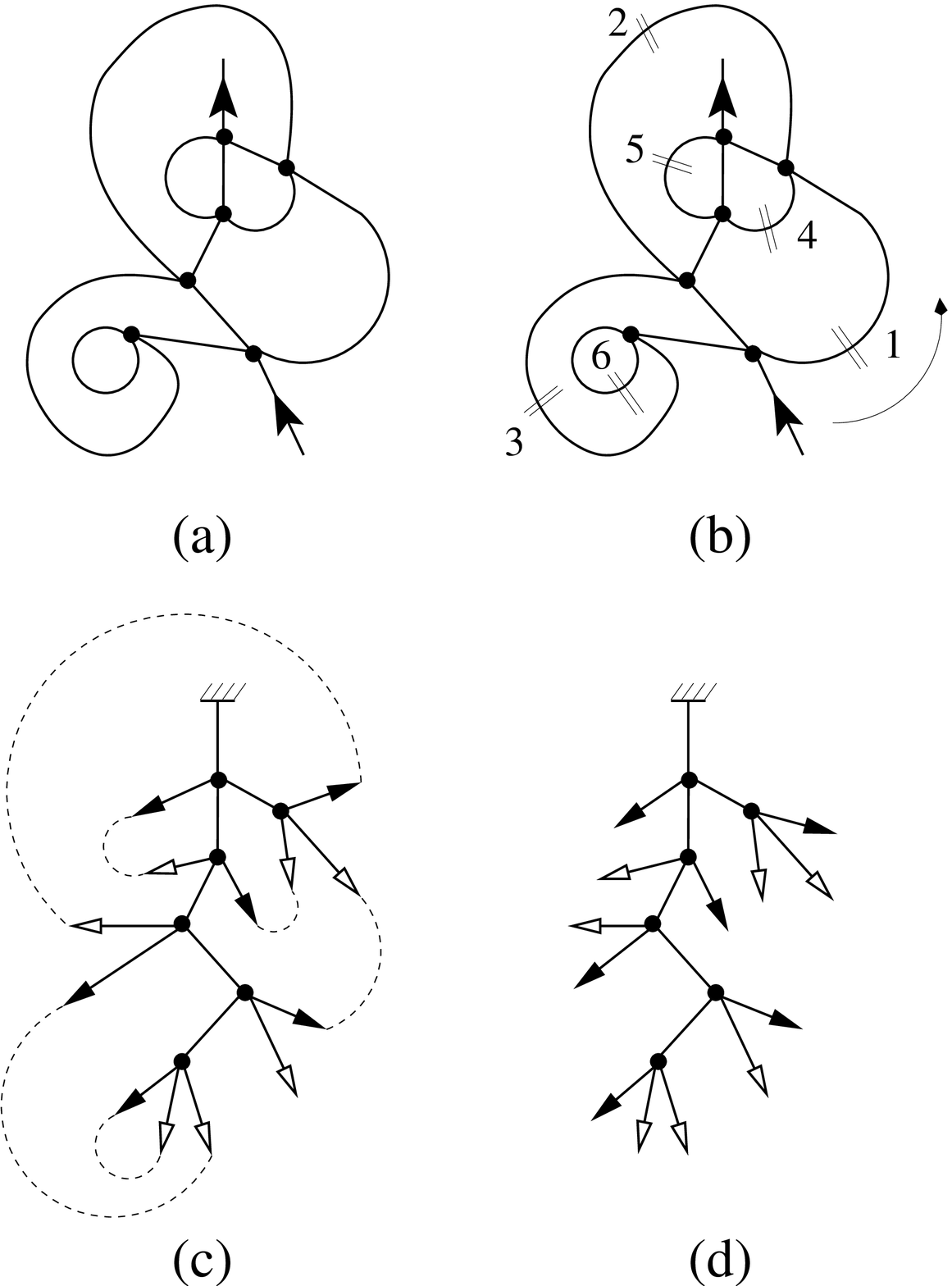}{8.cm}
\figlabel\bijecfour

For reasons of simplicity, let us start with the case of 4-valent 
graphs. Given a two-leg such graph $G$ (see Fig.\bijecfour\ for an illustration), 
we represent it in the plane by picking the 
external face to be adjacent to the first (in-coming) leg. We now visit all edges bordering this
external face in counterclockwise direction, and cut them iff the resulting graph remains connected.
We then replace the two halves of the cut edges by respectively a black and a white leaf.
This ``first passage" has merged a number of faces of the initial graph with the external one.
We now repeat the algorithm with the new external face, and so on until all faces are merged.
The resulting graph is a 4-valent tree $T$ (by construction, it has only one face and is connected).
The tree is then rooted at its second (outcoming) leg, while its incoming one is replaced 
with a white leaf.
Attaching a charge $+1$ (resp. $-1$) to white (resp. black) leaves, we obtain a tree with
total charge $+1$. It is easy to convince oneself that the resulting 4-valent tree has exactly
one black leaf at each vertex. 
\fig{The only two possibilities for the environment of an edge in a 4-valent blossom-tree,
obtained by cutting a two-leg planar 4-valent graph.
The edge separates the tree into a top and a bottom piece.
The first leg of the graph is chosen to be in the bottom piece.
The two cases correspond to whether the cutting process
stops in the top (a) or bottom (b) piece.
We have represented in both cases only the leaves unmatched within each piece.
In each case, the position of the root (second leg) is fixed by the fact that any 4-valent tree
must have an even number of leaves (including the root). We have indicated the 
corresponding charges $q=0$ or $+1$ of the top and bottom pieces.
}{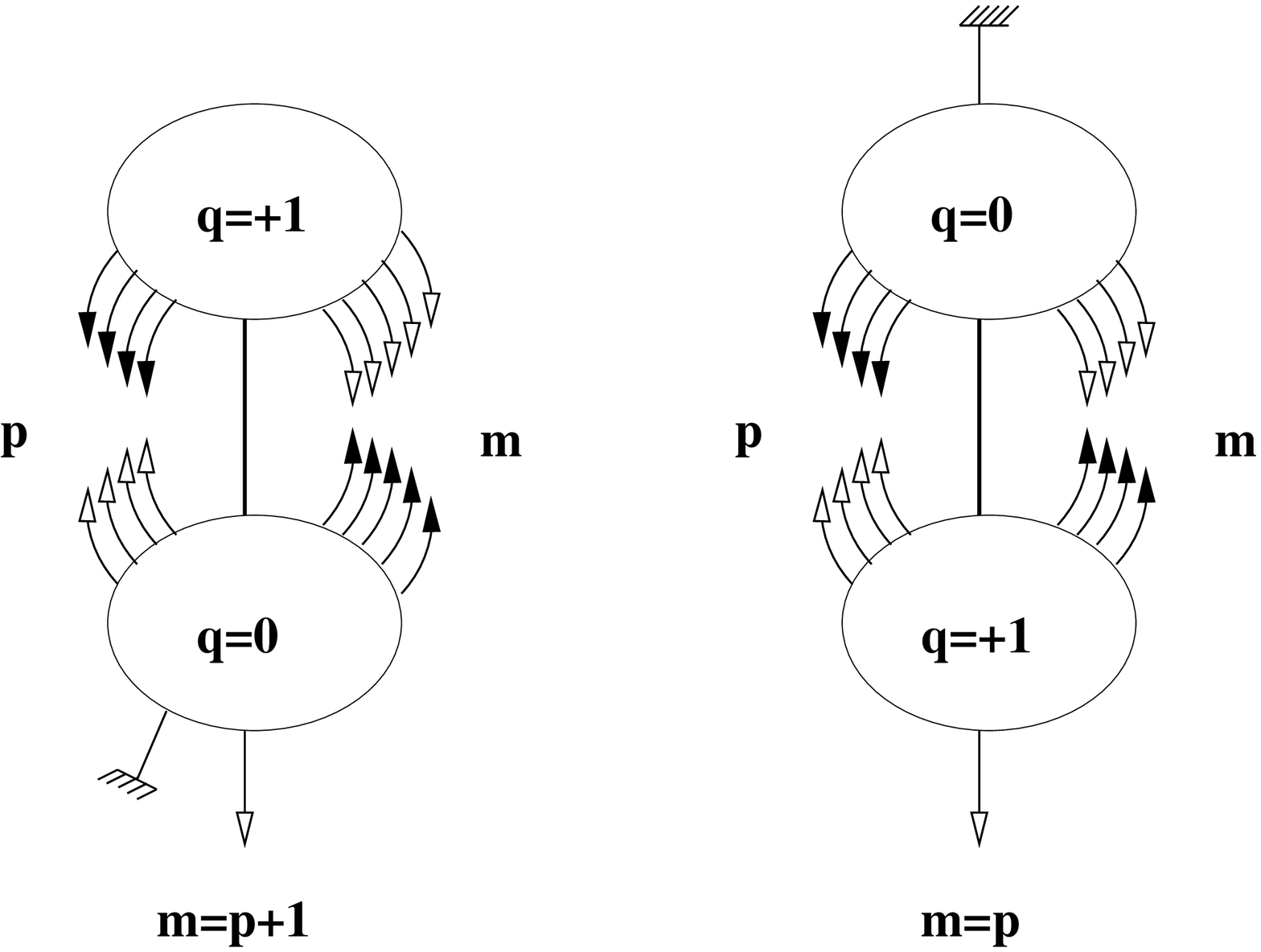}{9.cm}
\figlabel\onepervert
\noindent This is best proved by showing
that its descendent subtrees not reduced to a black leaf all have charge $+1$.
To see why, consider any edge of the blossom tree, not directly attached to a black leaf. 
It separates the tree into two (top and bottom)
pieces as depicted in Fig.\onepervert. As a result of the above iterative cutting procedure, we may keep
track of the $m$ and $p$ cut edges encompassing this edge, respectively lying on its right
and left, and connecting the top and bottom pieces. Assuming the first leg was in the bottom part,
and as the cutting process travels in counterclockwise direction, we may only have $m=p+1$
or $m=p$ according to whether the cutting process stopped in the top or bottom piece. But as the
top and bottom pieces are trees with only 4-valent inner vertices, they must have an even number 
of leaves, including the root, and the cut edge. 
Eliminating those matched by black/white pairs within each piece, we are respectively
left with: in case (a),
$2p+2$ leaves on top and $2p+3$ on the bottom, hence the root must be in the bottom; in case (b),
$2p+1$ leaves on top and $2p+2$ on the bottom, hence the root must be on top.
Adding up the charges, we see that the descendent piece (not containing the root) always has
charge $q=+1$.

Let us now define {\it rooted blossom-trees} as rooted
planar 4-valent trees with black and white leaves, a total charge $+1$, and exactly one black leaf
at each vertex (or equivalently such that each subtree not reduced to a black leaf has charge $+1$). 
Then the rooted blossom-trees are in bijection with the two-leg 4-valent 
planar graphs. The inverse mapping goes as follows. Starting from a rooted blossom-tree $T$,
we build a two-leg 4-valent planar graph by
connecting in counterclockwise direction around the tree all pairs of black/white leaves immediately 
following one-another, and by repeating this until all black leaves are exhausted. This leaves us 
with one unmatched white leaf, which we replace by the first leg, while the root becomes
the second leg. The order in which leaves are connected exactly matches the inverse of
that of the above cutting procedure.
This bijection now allows for a direct and simple counting of 2-leg 4-valent planar graphs, as
we simply have to count rooted blossom-trees.
Decomposing such trees according to the environment of the first vertex attached to their
root, we get the following equation for their generating function
\eqn\funcgenqua{
R=\quad 1 \qquad + \qquad \qquad 3 g R^2 }
$$\figbox{10.cm}{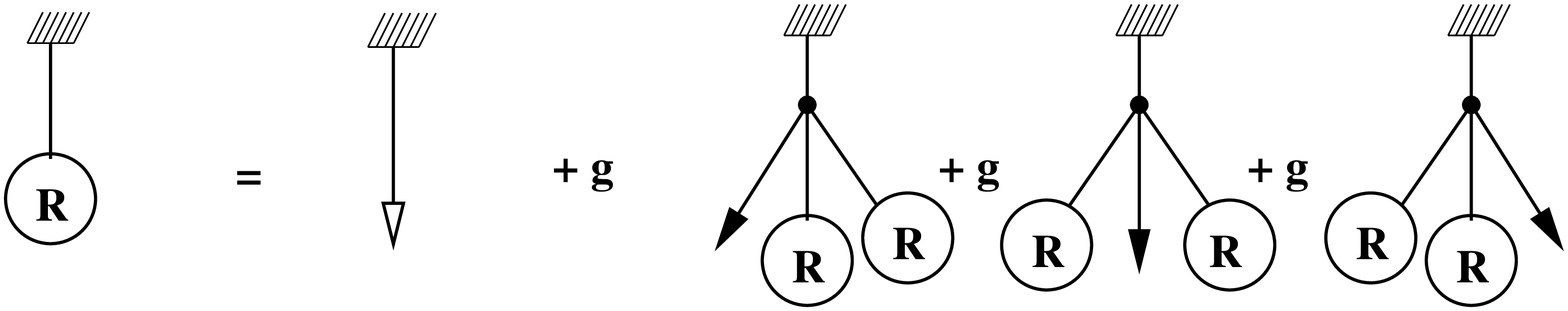}$$
where the first term corresponds to no vertex (and a white leaf directly connected to the root),
and the three others to a vertex with one black leaf and two descendent blossom-trees, each receiving
a weight $g$ for the decomposed vertex. Note that eq.\funcgenqua\ trivially amounts 
to the first equation of \quareduce. We have therefore found a purely combinatorial re-derivation
of the one-cut large $N$ matrix model result for planar 4-valent graphs, which confirms
its validity.

\subsec{Generalizations}

More generally, the above bijection may be adapted to two-leg planar graphs with arbitrary even 
vertex valences. Repeating the above cutting procedure on such a
two-leg planar graph leaves us with a rooted tree with 
only even vertex valences, with black and white leaves, and a total charge $+1$,
but with now exactly $k-1$ black leaves attached to each of its $2k$-valent vertices.
This is again a consequence of the equivalent property that any subtree not reduced to a black
leaf has charge $+1$, a fact proved exactly in the same manner as  before (actually, Fig.\onepervert\
is still valid for the case of arbitrary even valences).
This suggests a straightforward generalization of rooted blossom-trees with arbitrary
even vertex valences, with black and white leaves and such that any subtree not reduced to a black
leaf has charge $+1$. The latter are again in bijection with the two-leg planar graphs
with even valences, and are easily enumerated by considering the environment
of the vertex attached to the root, with the result for the generating function
$R$, including weights $g_{2k}$ per $2k$-valent vertex:
\eqn\Reveneq{R=1+\sum_{k\geq 1} g_{2k} {2k-1\choose k} R^k }
where the first term corresponds as in \funcgenqua\ to the tree with no vertex, while the $k$-th
term in the sum corresponds to the ${2k-1\choose k}$ ways of picking the $k-1$ black leaves
among the $2k-1$ descendents of the $2k$-valent vertex attached to the root, the remaining descendents
being themselves trees of charge $+1$ generated by $R$. 
The equation \Reveneq\ is nothing but \onlyone, written in a different fashion.

Finally, the bijection may be adapted so as to also include arbitrary (both even or odd)
valences, but then requires the introduction of one-leg graphs as well. Such graphs are represented
in the plane with their unique leg not necessarily adjacent to the external face, hence
are not generated by $\Gamma_1=S+V'_{-2}$, but, as we showed in the previous section, by $S$ itself.
The graphs are again cut according to the above procedure, to produce rooted trees.
The system of equations \rssystem\ is nothing but that obeyed by the rooted blossom-trees of two kinds
corresponding to cutting one- and two-leg graphs,
respectively generated by $S$ and $R$, and defined as rooted trees with black and white leaves,
and total charge $0$ and $+1$ respectively, and whose descendent subtrees not reduced to a black
leaf all have charge $0$ or $+1$. A simple way of recovering all combinatorial factors
in the two lines of \rssystem\ is to note that in a rooted blossom tree of charge $0$ 
(resp. $+1$), the $i-1$
descendents subtrees of any $i$-valent vertex attached to the root may be either black leaves (charge $-1$),
blossom trees of charge $0$, or blossom trees of charge $1$, the total charge being $0$ (resp. $+1$).
These subtrees are generated respectively by the functions $1$, $S$ and $R$.
Denoting by $j$ in both cases the total number of descendent subtrees of charge $+1$, we must have
$j$ (resp. $j-1$) black leaves to ensure the correct total charge, and the remaining $i-2j-1$ (resp.
$i-2j$) descendents have charge $0$. The combinatorial factors of \rssystem\ account for the
possible choices of these among the $i-1$ descendents.

This combinatorial interpretation sheds light on the algebraicity of the equations obtained 
in the large $N$ limit for the general one-matrix model: trees are indeed archetypical objects
whose generating functions obey algebraic relations, and we have shown that the planar graphs
generated by the large $N$ matrix model could be represented by (blossom) trees.
This correspondence will be fully exploited in Sect.6 to investigate the intrinsic 
geometry of planar graphs.

\newsec{The one-matrix model II: topological expansions and quantum gravity}

We now turn to higher genus contributions to the one-matrix model free energy.
This is best done by use of the so-called orthogonal polynomial technique \ORTHO.

\subsec{Orthogonal polynomials}

The standard technique of computation of \obtz\ uses orthogonal polynomials.
The idea is to disentangle the Vandermonde determinant squared interaction
between the eigenvalues.
The solution is based on the following simple lemma: 
if $p_m(x)=x^m+\sum_{j=0}^{m-1} p_{m,j}x^j$
are monic polynomials of degree $m$, for $m=0,1,...,N-1$, then
\eqn\vandereq{ \Delta(m)=\det(m_i^{j-1})_{1\leq i,j \leq N} =
\det(p_{j-1}(m_i))_{1\leq i,j \leq N} }
easily derived by performing suitable linear combinations of columns.
Let us now introduce the unique set of monic polynomials $p_m$, 
of degree $m=0,1,...,N-1$, that are
orthogonal w.r.t. the real one-dimensional
measure $d\mu(x)=\exp(-NV(x))dx$, namely such that
\eqn\monort{(p_m,p_n)= \int_{\IR} p_m(x)p_n(x) d\mu(x) = h_m \delta_{m,n} }
These allow us to rewrite the numerator of \obtz, using \vandereq,  as
\eqn\numera{ \sum_{\sigma,\tau \in S_N}
\epsilon(\sigma \tau) \prod_{i=1}^N \int_{\IR} d\mu(m_i) p_{\sigma(i)-1}(m_i)
p_{\tau(i)-1}(m_i) e^{-NV(m_i)} = N! \prod_{j=0}^{N-1} h_j }
We may apply the same recipee to compute the denominator, with the result
$N! \prod_{j=0}^{N-1} h_j^{(0)}$, where the $h_j^{(0)}$ are the squared norms
of the orthogonal polynomials w.r.t. the Gaussian measure $d\mu_0(x)=\exp(-Nx^2/2)dx$.
Hence the $h$'s determine $Z_N(V)$ entirely through
\eqn\thru{ Z_N(V)=\prod_{i=0}^{N-1} {h_i\over h_i^{(0)} } }
 
To further compute the $h$'s, let us introduce the two following operators $Q$ and
$P$, acting on the polynomials $p_m$:
\eqn\opq{ \eqalign{ Qp_m(x)&= x p_m(x) \cr
Pp_m(x) &= {d \over dx} p_m(x) \cr}}
with the obvious commutation relation 
\eqn\canocom{ [P,Q]=1}
Using the self-adjointness of $Q$ w.r.t.the scalar product $(f,g)=\int f(x)g(x) d\mu(x)$,
it is easy to prove that 
\eqn\qrec{ Qp_m(x)=xp_m(x)=p_{m+1}(x)+s_m p_m(x)+r_m p_{m-1}(x) }
for some constants $r_m$ and $s_m$,
and that $s_m=0$ if the potential $V(x)$ is even. 
The same reasoning yields
\eqn\rsm{ r_m={h_m \over h_{m-1}}, \ \ m=1,2,...} 
and we also set $r_0=h_0$ for convenience.
 
Moreover, expressing both $(Pp_m,p_m)$ and $(Pp_m,p_{m-1})$ in two ways, 
using integration by parts, we easily get the master equations
\eqn\masteq{\eqalign{ {m\over N} &= {(V'(Q)p_m,p_{m-1})\over (p_{m-1},p_{m-1})} \cr
0&=(V'(Q)p_m,p_m) \cr}}
which amount to a recursive system for $s_m$ and $r_m$. Note that the second line 
of \masteq\ is automatically satisfied if $V$ is even: it vanishes as the integral
over $\IR$ of an odd function. 
Assuming for simplicity that $V$ is even, the first equation of \masteq\ gives
a non-linear recursion relation for the $r$'s, while the second is a tautology, due to the
vanishing of all the $s$'s:
\eqn\evenrecr{\eqalign{ {m\over N}&= {(V'(Q)p_m,p_{m-1})\over (p_{m-1},p_{m-1})}=
\sum_{k\geq 1} g_{2k} {(Q^{2k-1} p_m,p_{m-1})\over (p_{m-1},p_{m-1})} \cr
&=\sum_{k\geq 1} g_{2k} \sum_{{\rm paths} \ p \vert p(1)=m,\ p(2k-1)=m-1\atop
p(i+1)-p(i)=\pm 1} \prod_{i=1}^{2k-2} w(p(i),p(i+1)) \cr}}
where the sum extends over the paths $p$ on the non-negative integer line, with $2k-1$
steps $\pm 1$, starting at $p(1)=m$ and ending at $p(2k-1)=m-1$, and the weight reads 
$w(p,q)=1$ if $q=p+1$, and $w(p,q)=r_p$ if $q=p-1$.
For up to 6-valent graphs this reads
\eqn\mainequpsix{\eqalign{
{n\over N}&= r_n(1-g_2) -g_4r_n(r_{n+1}+r_n+r_{n-1})\cr
&-g_6(r_{n+1}r_{n+2}+r_{n+1}r_{n-1}+
r_{n-1}r_{n-2}+r_n^2+r_{n+1}^2+r_{n-1}^2+2r_n(r_{n+1}+r_{n-1}) \cr} }
In general, the degree $d$ of $V$ fixes the 
number $d-1$ of terms in the recursion. So, we need to feed the $d-2$ initial values 
of $r_0,r_1,r_2,...,r_{d-3}$ into the recursion relation, and we obtain the exact value of
$Z_N(V)$ by substituting $h_i=r_0r_1...r_i$ in both the numerator and the denominator
of \thru. Note that for $V_0(x)=x^2/2$ the recursion \masteq\ reduces simply to
\eqn\zercas{ {m\over N}= {(Qp_m^{(0)},p_{m-1}^{(0)})\over (p_{m-1}^{(0)},p_{m-1}^{(0)})}
= r_m^{(0)} }
and therefore $h_m^{(0)}=h_0^{(0)} m!/N^m =\sqrt{2\pi} m!/N^{m+1/2}$. The $p_m^{(0)}$ are 
simply the (suitably normalized) Hermite polynomials.  

Finally, the full free energy of the model \onmav\ reads
\eqn\frenen{ F_N(V)={\rm Log}\, Z_N(V)=N\,{\rm Log}\,r_0\sqrt{N\over 2\pi}\, + 
\sum_{i=1}^{N-1} (N-i){\rm Log} {Nr_i\over i}} 
in terms of the $r$'s. 

\subsec{Large $N$ limit revisited}

In view of the expression \frenen, it is straightforward to get 
large $N$ asymptotics for the free
energy, by first noting that as $h_0\sim \sqrt{2\pi \over N}$,
the first term in \frenen\ doesn't contribute to the leading order $N^2$
and then by approximating the sum by an integral of the form
\eqn\intaprox{ f=\lim_{N\to \infty} {1\over N} \sum_{i=1}^{N-1}
(1-{i\over N}) {\rm Log} {r_i\over i/N} = \int_0^1 dz (1-z) {\rm Log}{r(z)\over z}}
where we have assumed that the sequence $r_i$ tends to a function $r_i\equiv r(i/N)$ of
the variable $z=i/N$ when $N$ becomes large. This assumption, wrong in general, 
basically amounts to the one-cut hypothesis encountered in Sect.3.2.
The limiting function $r(z)$ in \intaprox\ is then determined by the equations
\masteq, that become polynomial in this limit. In the case $V$ even for 
instance, we simply get
\eqn\req{ z= r(z)-\sum_{k\geq 1} {2k-1 \choose k} g_{2k} r(z)^k }
The function $r(z)$ is the unique root of this polynomial equation that tends to $z$
for small $z$ (it can be expressed using the Lagrange inversion method for instance,
as a formal power series of the $g$'s), and the free energy follows from \intaprox.  
To relate this expression to our former results, let us again attach an extra weight
$t$ per face of the graphs. As before, it amounts to replacing
$V'(x)\to V'(tx)=tx-\sum_{k\geq 1} g_{2k}t^{2k-1} {2k-1\choose k} x^{2k-1}$, and to rescale $f\to t^2 f$.
Setting $\rho(z)=t^2 r(z)$, we arrive at
\eqn\arriven{ tz= \rho(z)-\sum_{k\geq 1} {2k-1 \choose k} g_{2k} \rho(z)^k \equiv \varphi(\rho(z))}
and $f=\int_0^1 dz (1-z) {\rm Log}{\rho(z)\over tz}$, $\rho(z)$ being determined
by $\varphi(\rho(z))=t z$. Let us perform in this integral the change 
of variables $z\to \rho$, with
$dz=\varphi'(\rho)/t d\rho$, and integration bounds $\rho(0)=0$ and $\rho(1)=\Rho$, 
solution of $t=\varphi(\Rho)$. We obtain:
\eqn\chgvars{ t^2f=\int_0^\Rho d\rho \varphi'(\rho) 
(t-\varphi(\rho))\ {\rm Log}\,{\rho\over \varphi(\rho)}}
We now take derivatives w.r.t. $t$: as the dependence on $t$ is either via $\Rho$ or explicit in the integrand,
there are only two terms involved. But the integrand vanishes at the upper bound, as $t-\varphi(\Rho)=0$,hence
only the explicit derivative contributes, and we have
\eqn\derivaphi{ \eqalign{
\partial_t(t^2 f)&= \int_0^\Rho d\rho \varphi'(\rho) 
\ {\rm Log}\,{\rho\over \varphi(\rho)}\cr
\partial_t^2(t^2 f)&= \partial_t \Rho \varphi'(\Rho) 
\ {\rm Log}\,{\Rho\over \varphi(\Rho)}={\rm Log}\,{\Rho\over t}\cr}}
Note that $\Rho$ may be interpreted in the light of Sect.4 as the generating function for 
rooted blossom trees with a weight $t$ per white leaf (easily read off the relation
$\Rho=t+\sum_{k\geq 1} g_{2k}{2k-1\choose k} \Rho^k$). 
Finally, setting $t=1$, we may rewrite 
\eqn\rewfintf{\partial_t^2(t^2 f)\vert_{t=1}={\rm Log}\,R=
-{\rm Log}\big(1-\sum_{k\geq 1} g_{2k}{2k-1\choose k} R^{k-1}\big)}
as $\Rho$ reduces to $R$ at $t=1$. This expresses the generating function for planar graphs
with even valences and with two distinct marked faces (as each derivative amounts to a marking)
as the logarithm of the generating function for blossom trees. This formula will become 
combinatorially clear in Sect.7.2 below.

\subsec{Singularity structure and critical behavior}

In the even potential case, according to \derivaphi, the singularities of $\Rho$ govern
those of the free energy. $\Rho$ attains a first critical singularity at some $t=t_c$ where $\Rho=\Rho_c$
with $\varphi(\Rho_c)=t_c$ and $\varphi'(\Rho_c)=0$. We may then Taylor-expand
\eqn\talorex{ t_c-t=\varphi(\Rho_c)-\varphi(\Rho)=-{1\over 2}(\Rho-\Rho_c)^2\varphi''(\Rho_c)+
O((\Rho-\Rho_c)^3)}
As $t$ is an activity per face of the graphs, we may consider the number of faces as a measure of
the area of the associated discrete surface, therefore the singularity 
$\Rho_{sing}\sim (t_c-t)^{1/2}$ is immediately translated via \derivaphi\ into a singularity
of the planar free energy $f_{sing}\sim  (t_c-t)^{2-\gamma}$, with a string susceptibility
exponent $\gamma=-1/2$. Alternatively, upon Laplace-transforming the result, this exponent also 
governs the large area behavior of $f_A\sim {\rm const.}\ t_c^{-A}/A^{3-\gamma}$, 
the planar free energy for fixed area ($A$=number of faces here).
This is the generic singularity expected from a model 
describing space without matter, such as that of the pure 4-valent graphs studied above.

We may reach more interesting multicritical points with
different universality classes and exponents by fine-tuning the parameters 
$g_{2k}$ so as to ensure that a higher order singularity is attained at some $t=t_c$ such that
$\Rho=\Rho_c$, while $\varphi'(\Rho_c)=\varphi''(\Rho_c)=...=\varphi^{(m)}(\Rho_c)=0$,
while $\varphi^{(m+1)}(\Rho_c)\neq 0$. Taylor-expanding now yields
\eqn\talom{ t_c-t=\varphi(\Rho_c)-\varphi(\Rho)=
-{\varphi^{(m+1)}(\Rho_c)\over (m+1)!}(\Rho-\Rho_c)^{m+1}+O((\Rho-\Rho_c)^{m+2})}
This translates into a singularity of the free energy with string susceptibility
exponent $\gamma=-{1\over m+1}$. This is characteristic of non-unitary matter conformal field theory
with central charge $c(2,2m+1)$ coupled 
to 2D quantum gravity \CFT\ \KPZ. The first example of this is the Hard Dimer model introduced in Sect.2.4
above, for which
\eqn\phihd{\varphi_{HD}(\Rho)=\Rho-3g\Rho^2-30 z g^2 \Rho^3}
Writing $\varphi_{HD}'(\Rho)=\varphi_{HD}''(\Rho)=0$ yields $z_c=-1/10$, $g\Rho_c=1/3$, and
$gt_c=1/3$, with a critical exponent $\gamma=-1/3$, corresponding to the Lee-Yang edge singularity
(conformal field theory with central charge $c(2,5)=-22/5$) coupled to 2D quantum gravity.

The inclusion of vertices of odd valences does not give any additional multicritical singularities.
This is why we choose to stick here and in the following to the even case as much as possible.

\subsec{Higher genus}

To keep the full fledge of the model, we must keep track of all shifts of indices in \evenrecr.
This is easily done by still introducing $r(z=m/N)\equiv r_m$, but by also keeping
track of finite shifts of the index $m\to m+a$, namely, setting $\epsilon=1/N$, via
$r(z+a \epsilon)\equiv r_{m+a}$. In other words, as $N\to \infty$, we still assume
that $r_m$ becomes a smooth function of $z=m/N$, but keep track of finite index
shifts. Solving eq.\evenrecr\ order by order in $1/N$ involves writing the ``genus" expansion
\eqn\genusr{ r(z)=\sum_{k\geq 0} \epsilon^{2k} r^{(k)}(z), }
implementing all finite index shifts by the corresponding $\epsilon$ shifts of the
variable $z$, and solving for the $r^{(k)}$'s order by order in $\epsilon^2$. 
We finally have to substitute the solution back into the free energy \frenen, with $r_i=r(i/N)$.
This latter expression must then be expanded order by order in $\epsilon$ using the 
Euler-MacLaurin formula. Setting $h(x)=(1-x){\rm Log}(r(x)/x)$, this gives
\eqn\givfn{
{F_N(V)\over N^2}={1\over N}\sum_{i=1}^N h\left({i\over N}\right)=\int_0^1 h(z)dz
+{\epsilon\over 2}(h(1)-h(0))+{\epsilon^2\over 12}(h'(1)-h'(0))+...}
in which we must also expand $r(x)$ according to \genusr.
The result is the genus expansion $F_N(V)=\sum N^{2-2h} F^{(h)}(V)$, where $F^{(h)}$ is
the generating function for graphs of genus $h$.
For illustration, in the 4-valent case, we have
\eqn\matequa{ r(z)-gr(z)(r(z+\epsilon)+r(z)+r(z-\epsilon))=z}
Writing $r(z)=r^{(0)}(z)+\epsilon^2 r^{(1)}(z)+O(\epsilon^4)$, we find that
\eqn\genmusone{ r^{(1)}(z)(1-6 gr^{(0)}(z))=gr^{(0)}(z){r^{(0)}}''(z) }
at order $2$ in $\epsilon$, while $r^{(0)}(z)=(1-\sqrt{1-12 gz})/(6g)$, and $(1-6 gr^{(0)}(z)){r^{(0)}}'(z)=1$,
so that $r^{(1)}(z)=gr^{(0)}(z){r^{(0)}}'(z){r^{(0)}}''(z)$.
At to order $2$ in $\epsilon$, this gives 
\eqn\quagenone{ F^{(1)}= {1\over 12}(h_0'(1)-h_0'(0)) +g \int_0^1 dx(1-x) {r^{(0)}}'(x){r^{(0)}}''(x)}
where $h_0(x)=(1-x){\rm Log}(r^{(0)}(x)/x)$, namely
\eqn\expanfone{\eqalign{ F^{(1)}&={1\over 24}\sum_{n\geq 1} {g^n\over n}3^n(4^n-{2n\choose n})\cr
&={g\over 4}+\qquad\qquad {15\over 8} g^2+...\cr}}
$$\figbox{6.cm}{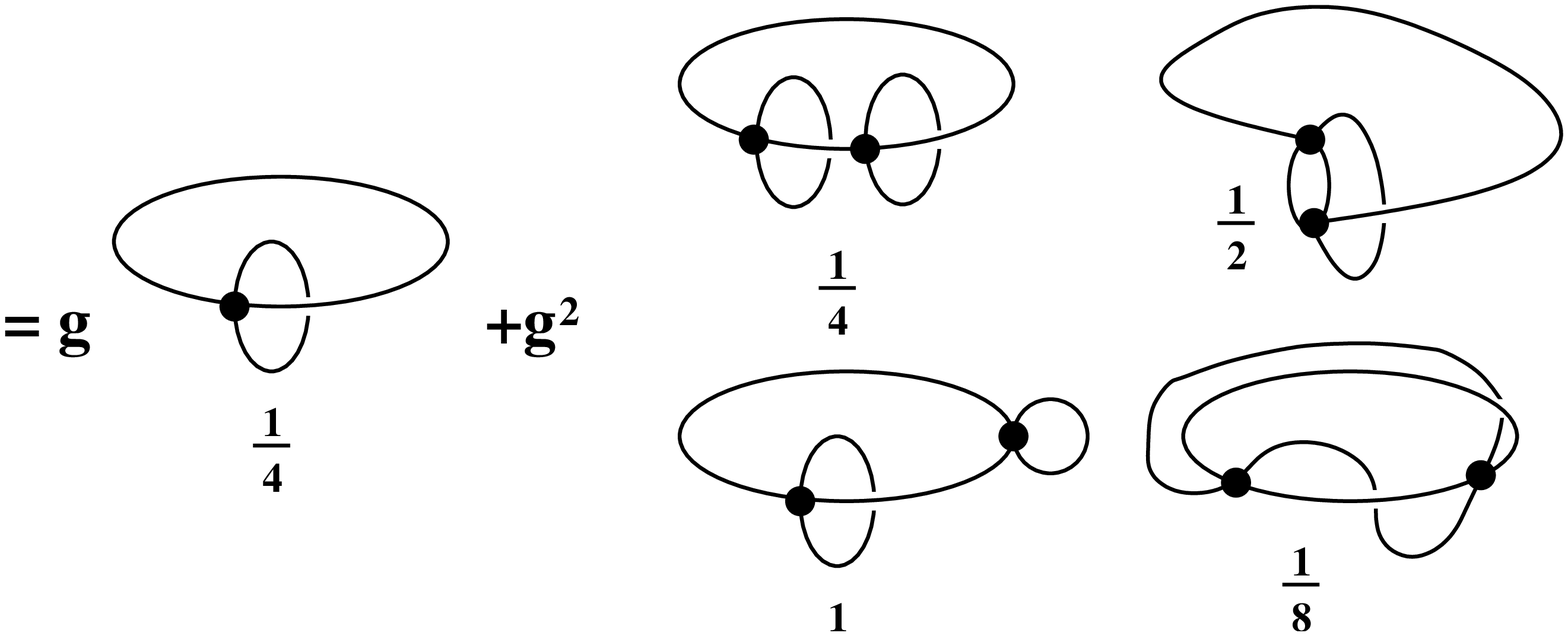}$$
where we have displayed the genus one 4-valent graphs with up to two vertices,
together with their inverse symmetry factors.

\subsec{Double-scaling limit}

The idea behind the double-scaling limit is to combine the large $N$ limit and the singularity
structure of the free energy at all genera into a single scaling function. Let us first consider the
4-valent case \matequa. We wish to approach the critical value $g=g_c=1/12$ displayed by the 
planar solution $R=(1-\sqrt{1-12g})/(6g)$ at $t=1$, {\it at the same time} as $N\to\infty$.
Setting $\rho(z)=gr(z)$, we have 
\eqn\gzt{\eqalign{ gz&=\rho(z)(1-(\rho(z+\epsilon)+\rho(z)+\rho(z-\epsilon)))\cr
g_c&=\rho_c(1-3\rho_c)\cr}}
with $g_c=1/12$ and $\rho_c=1/6$. 
Subtracting both lines of \gzt\ and expanding up to order $2$ in $\epsilon$ yields
\eqn\epantwogtz{g_c-g z= \rho_c(1-3\rho_c)-\rho(z)(1-3\rho(z))-\epsilon^2\rho(z)\rho''(z)+O(\epsilon^4)}
This suggests to introduce rescaled
variables and functions $g_c-g z=a^2 g_c y$, while $\rho(z)=\rho_c(1-a u(y))$, and to expand
up to order $2$ in $a$ as well:
\eqn\epantwogtza{a^2 g_c y=3\rho_c^2 a^2 u(y)^2-\epsilon^2\rho_c^2 a^{-3}u''(y)}
where we have noted that $dz= -a^2 dy$ at $g=g_c$. The large $N$ limit of Sect.5.2 is recovered
by taking $\epsilon=0$, in which case we are left with $u(y)=\sqrt{y}$, another way of expressing
the planar singularity of the free energy $u(y)=y^{-\gamma}$, with $\gamma=-1/2$.
For non-zero $\epsilon$, all terms in \epantwogtza\ will contribute if we take $\epsilon^2= a^5$.
We then have 
\eqn\pleve{ y=u(y)^2-{1\over 3} u''(y) }
which is nothing but the Painlev\'e I equation.
Moreover, the singular part of the free energy reads
\eqn\frenera{F\equiv F_{sing}=N^2\int_0^1 dz (1-z){\rm Log}({\rho(z)\over gz})\vert_{sing}
=N^2a^5 \int_{a^{-2}}^x(y-x)u(y) dy}
where $g_c-g=a^2 g_c x$. Differentiating twice w.r.t. $x$ yields $u(x)=-F''(x)$.
To summarize, if we take simultaneously $N\to \infty$ and $g\to g_c$, but keep the quantity
\eqn\dsl{ N^{4\over 5} \left({g_c-g\over g_c}\right)= x} 
fixed, then the singular parts of the free energy at all genera recombine into a single 
scaling function $F(x)$, whose second derivative satisfies the Painlev\'e I differential equation.
To recover the leading singularity at genus $h$, we simply have to expand
the solution of \pleve\ at large $x$ as $u(x)=\sum_{h\geq 0} u_h x^{{1\over 2}(1-5h)}$
and solve the resulting recursion relation for $u_h$.
This is the so-called double-scaling limit of pure 2D quantum gravity.

We may repeat this exercise with the multicritical models of Sect.5.3, however algebra becomes
cumbersome. Let us instead look at the scaling limits of the operators $P$ and $Q$ acting on
the orthogonal polynomials \opq. 
Let us rescale the orthogonal polynomials $p_n$ to make them
orthonormal, namely set ${\tilde p}_n=p_n/\sqrt{h_n}$, so that
\qrec\ (with $s_n=0$) becomes more symmetric
\eqn\recnewq{ (Q{\tilde p})_n~=~ \lambda {\tilde p}_n~=~
\sqrt{r_{n+1}} {\tilde p}_{n+1}+\sqrt{r_n} {\tilde p}_{n-1}}
or equivalently
\eqn\eqiq{ Q_{n,m}~=~({\tilde p}_m,Q{\tilde p}_n)~=~\sqrt{r_{n+1}}
\delta_{m,n+1}+ \sqrt{r_n} \delta_{m,n-1} }
Let us now take the large $N$ limit. Setting $\epsilon=1/N$ as before,
we note that the shift operator $\delta_{m,n+1}$, acting on sequences 
$(\alpha_m)$ can
be generated as $e^{\epsilon d/dz}$, acting on the continuum limit of 
$(\alpha_m)$, i.e. a function $\alpha(z)=\alpha_m$, for $z=m/N$. Indeed, one just
has to write
\eqn\juste{ \sum_m \delta_{m,n+1} \alpha_m ~=~\alpha_{n+1}~
=~e^{d \over dn} \alpha_n\simeq
e^{\epsilon {d \over dz}} \alpha(z)}
Setting again $r(z)=r_c(1-au(y))$, this permits to rewrite $Q$ as
\eqn\limiqc{\eqalign{
Q~&\simeq~ \sqrt{r(z)} \left( e^{\epsilon {d \over dz}}+
e^{-\epsilon {d \over dz}}\right)\cr
&=~\sqrt{{r_c}(1-a u(y))} ( 2 +\epsilon^2 {d^2 \over dz^2}+O(\epsilon^4))\cr
&=~2\sqrt{r_c} -\sqrt{r_c}(au-(\epsilon {d\over dz})^2 +O(\epsilon^4,a^2))\cr}}
In the general multicritical case, we must set
\eqn\setmulti{ t_c-tz=a^{m+1} t_c y}
and $y=x$ at $z=1$, so that $dz\sim a^{m+1} dy$.
The two terms in the r.h.s. of \limiqc\ are of the same order $a$ provided
$(\epsilon d/dz)^2=(\epsilon a^{-m-1} d/dy)^2$ is of order $a$, and we obtain
the double-scaling condition that $\epsilon^2a^{-2m-2}=a$, hence 
\eqn\valndsl{ N^2=a^{-2m-3} }
or equivalently
\eqn\dslm{ N^{2m+2\over 2m+3} \left( {t_c-t\over t_c}\right)=x}
remains fixed while $N\to\infty$ and $t\to t_c$.
Retaining only the
coefficient of $a$, we find that
\eqn\limiofq{ Q \to {d^2\over dy^2} -u(y) }
in the double-scaling limit. This limit is a differential operator,
acting on functions of the rescaled variable $y$.

Let us now turn to $P$. It will be useful to slightly change the 
definition of the operator $P$, in the following way
\eqn\defpmo{\eqalign{ P_{n,m}~
&=~ \int_{-\infty}^\infty d\lambda
{\tilde p}_m(\lambda) e^{-N{V(\lambda) \over 2}} {d\over d\lambda}
e^{-N{V(\lambda)\over 2}} {\tilde p}_n(\lambda)\cr
&=~-{N\over 2}({\tilde p}_m,V'(Q) {\tilde p}_n) +
({\tilde p}_m,{\tilde p}_n')\cr 
&=~-{N\over 2}V'(Q)_{n,m} +A_{n,m} \cr}}
where $A$ is a lower triangular matrix $A_{nm}=0$ if $n\leq m$.
Upon an integration by parts we may as well write
\eqn\aswell{ \eqalign{
P_{n,m}~&=~-{N\over 2}V'(Q)_{n,m} +A_{n,m}\cr 
&=~{N\over 2}V'(Q)_{n,m} -A^t_{n,m}\cr}}
where the matrix $A^t$ is upper triangular. Eq.\aswell\ permits
to compute the matrix elements of $P$ in terms of those of $Q$ only,
by using the first equation when $n\leq m$ ($A_{nm}=0$) and the second
one when $n\geq m$ ($A^t_{nm}=0$).
This can be summarized by the following operator relation:
\eqn\relop{ P~=~{N\over 2}(V'(Q)_+-V'(Q)_-)}
where the index $+$ (resp $-$) indicates that we retain only the upper (resp. lower)
triangular part.
In particular, as it is expressed polynomially in terms of $Q$,
$P$ has a finite range, namely $P_{n,m}=0$ if $|n-m|>B$, $B$ some
uniform bound, independent of $N$ ($B$ depends only 
on the degree of $V$). This bound ensures that $P$ goes over in
the double scaling limit to a differential operator of finite
degree $p$, of the form
\eqn\lipn{ P ={1 \over a\sqrt{r_c}} (d^p+ v_2 d^{p-2}+v_3 d^{p-3}+....+v_p) }
to ensure the correct normalization of $[P,Q]=1$.
From the precise form of $P$ \relop, and as each derivative $d$ w.r.t. $y$ carries
a prefactor $\epsilon a^{-m-1}=\sqrt{a}$, we must have $N\times a^{p/2}=1/a$,
which together with the double-scaling condition \valndsl\ fixes the degree
\eqn\degP{ {\rm deg}(P)= 2m+1}
We must finally write the canonical commutation relation \canocom\
$[P,Q]=1$, with the renormalized values $P=d^{2m-1}+ v_2 d^{2m-2}+....+v_{2m}$
and $Q=(d^2-u)$. 
Let us introduce the square root $L$ of $Q$, namely the unique pseudo-differential operator
\eqn\pseudo{ L=d+\sum_{i\geq 1} \ell_i d^{-i}}
such that $L^2=Q$. This equation is expressed as a triangular system for the $\ell$'s,
provided we normal-order the result by pushing all functions to the left 
of powers of the differential $d$, by means of the Leibnitz formula 
$d^{-i}f(y)=\sum_{j\geq 0} (-1)^j{i+j-1\choose j} f^{(j)}(y) d^{-i-j}$. Let us now express $P$.
Solving $[{\tilde P},Q]=0$ rather than $[P,Q]=1$ makes no difference as far as we only write
the equations for the coefficients of positive powers of $d$: solving these equations
precisely allows to express $P$ as a function of $Q$. As the solution to $[{\tilde P},Q]=0$
for a pseudo-differential operator ${\tilde P}$ of degree $2m+1$ is nothing but 
${\tilde P}=L^{2m+1}$, we simply have $P=(L^{2m+1})_+$, where the subscript $+$ indicates that
we have retained only the differential polynomial part. So far, we have solved all the equations obtained
by setting to $0$ the coefficients of all positive powers of $d$ in $[P,Q]=1$. We still have
to write the $d^0$ coefficient.
Writing $(L^{2m+1})_-=L^{2m+1}-(L^{2m+1})_+=R_{m+1}[u] d^{-1}+O(d^{-2})$, this last
equation reads simply
\eqn\redsimp{ 2 R_{m+1}[u]'=1\qquad  \Rightarrow \qquad  2 R_{m+1}[u]=y}
This is nothing but a higher order generalizations of the Painlev\'e I equation,
related to the so-called KdV hierarchy. From their definition, the ``KdV residues" 
$R_{m}[u]$ satisfy the recursion relation
\eqn\recukdv{ R_{m+1}[u]'= {1\over 4}R_m[u]'''-{1\over 2}u'R_m[u]-uR_m[u]'}
obtained by writing $(L^{2m+1})_-=((L^{2m-1})_-Q)_-=(Q(L^{2m-1})_-)_-$, while the initial term
reads $R_1[u]=-u/2$. Again, plugging the large $y$ expansion 
$u(y)=\sum_{h\geq 0} u_h y^{{1\over m+1}(1-(2m+3)h)}$ into eq.\redsimp\ yields a recursion relation
for the $u_h$ and gives acces to the all genus singular part of the free energy via
the relation $F_{sing}(x)''=-u(x)$.

The actual general solution of $[P,Q]=1$ involves integration constants which we have all set to
zero for convenience, hence the most general solution for a degree $2m+1$ differential operator $P$
reads
\eqn\trueq{2\sum_{j=1}^{m+1} \mu_j R_j[u]=y }
for some integration constants $\mu_j$. This equation interpolates between the various matter critical
points $\mu_j=\delta_{j,k+1}$, corresponding to the various multicritical points already identified
as $c(2,2k+1)$ CFT coupled to 2D quantum gravity. From the point of view of the $m+1$-critical model,
the $\mu$'s are just dimensionful parameters coupled to the order parameters of the theory.

\subsec{Generalization to multi-matrix models}

A large class of multi-matrix models turns out to be solvable by exactly the same techniques
as those developed in the previous sections for the one-matrix model. It corresponds
to matrices $M_1,...,M_p$ with a chain-like interaction, namely involving a quadratic form $Q_{a,b}$
as in \fromult, for which only the elements
$Q_{a,a}$, $a=1,2,...,p$ and say $Q_{a,a+1}$, $a=1,2,...,p-1$ are non-vanishing.
In this particular case only, the unitary group integrations may be disentangled from the
eigenvalue integrations for all $M$'s and we may still reduce the integral to one over
eigenvalues of the different matrices. Once this step is performed, the orthogonal 
polynomial technique is easily adapted and a complete solution follows from
considering again operators $P_a$ and $Q_a$ of differentiation w.r.t. or multiplication by
an eigenvalue of the matrix $M_a$, $a=1,2,...,p$. Note that the saddle-point technique
with several matrices is more subtle.

One is eventually left with
solving an equation of the form $[P_1,Q_1]=1$, the scaling function $u$ such that $u''=-F$
being identified with some coefficient of $Q_1$.  The remarkable fact is that both $P_1$ and $Q_1$
remain of uniformly bounded range, the latter depending only on the degrees of the potentials
for the various matrices. This implies that in a suitable double scaling limit where the size of the
matrices is sent to infinity and the parameters of the potentials
go to some (multi-) critical values,
the operators $P_1$ and $Q_1$ still become differential operators of finite degree
say $p$ and $q$, two coprime integers. The resulting differential system $[P_1,Q_1]=1$
governs the all-genus singular part of the free energy of the general $c(p,q)$ minimal conformal field
theory coupled to 2D quantum gravity. This completes the picture of critical behaviors covered
by matrix models solvable by orthogonal polynomial techniques: it exhausts 
all minimal CFT's with $c<1$, according to the famous ADE classification 
thereof \CFT\foot{This statement is not
completely correct: only the A-type CFT's are covered by the standard
multi-matrix models. A proposal for D-type CFT's was given in \DK, based on D-type
generalizations of the KP hierarchy \DS, but no direct relation to solvable
matrix models was found. Not to speak about E-type solutions...}.

{}From a combinatorial point of view, the bijection presented in Sect.4.1 may be generalized
to the case of two-matrix models \BMS, and presumably to all cases solvable by orthogonal
polynomial techniques, which all lead to algebraic systems, henceforth suggest tree-like
interpretations.

\newsec{The combinatorics beyond matrix models: geodesic distance in planar graphs}

In this section, we return to the bijections between planar graphs and trees to investigate 
more refined properties of the discrete random surfaces generated by matrix models, 
involving their intrinsic geometry. 
In particular, we will derive in a purely combinatorial manner
sets of closed equations for generating functions of planar graphs
with marked points at a given geodesic distance, a task still eluding the
matrix model description.

\subsec{Keeping track of the geodesic distance: the 4-valent case}

Let us return to the bijection between two-leg 4-valent planar graphs and rooted blossom trees
shown in Sect.4. Looking at Fig.\bijecfour, we see that the bijection allows to
keep track of the geodesic distance between the two legs, namely the smallest possible
number of edges of the graph crossed by a curve joining them. Indeed, this distance is nothing
but the number of edges encompassing the root of the corresponding blossom-tree, when black and
white leaves are re-connected. Loosely speaking, the geodesic distance between the legs
corresponds in the blossom tree language to the number of black leaves ``in excess", which require
encompassing the root to be connected to their white alter ego in counterclockwise direction.
Let us now derive simple relations for the generating function $R_n$ for two-leg diagrams with
geodesic distance at most $n$ between the legs.
To get the more interseting generating function $G_n$ for graphs with two legs at geodesic distance 
equal to $n$, we just have to write $G_n=R_n-R_{n-1}$.
Alternatively, $R_n$ can be thought of as the generating function for blossom trees
with at most $n$ black leaves in excess. As such, it obeys the following recursion relation:
\eqn\recurn{R_n=\quad 1\qquad +\qquad g R_{n+1}R_n \qquad+\qquad
g R_n^2\qquad+\ \ \qquad gR_nR_{n-1}}
$$\figbox{11.cm}{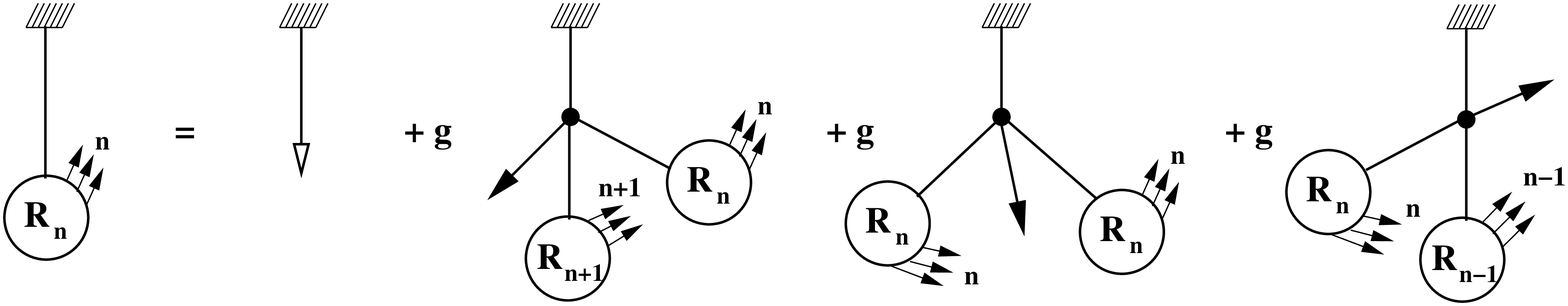}$$
This is just a refinement of eq.\funcgenqua\ in which we have kept track of the maximal
numbers of excess black leaves. The presence of single black leaves around the vertex
connected to the root lowers by $1$
the maximal number of excess leaves of any object on its left, while as 
each blossom tree has one white leaf in excess, it always absorbs one excess black 
leaf from objects on its left: these two facts are responsible for the shifts
of the index $n$. 

The recursion relation \recurn\ holds for all $n\geq 0$ provided the term involving $R_{-1}$
is dropped. Let us therefore supplement the recursion relation with the initial 
value $R_{-1}=0$. Moreover, the function $R_n$ should go over to the function $R$ of
\funcgenqua\ in the limit $n\to \infty$, which amounts to suppressing the constraint on the
distance between the two legs. $R_n$ is the unique solution to \recurn\ such that 
$R_{-1}=0$ and $\lim_{n\to \infty} R_n=R$. If we are only interested in the power series expansion 
of $R_n$ in $g$, we may solve \recurn\ order by order in $g$, starting with $R_{-1}=0$
at all orders and $R_n=1+O(g)$ for all $n\geq 0$. To any given order in $g$, the system for the
series coefficients is indeed triangular, and moreover $R-R_n=O(g^{n+1})$, which guarantees the
convergence condition.
In the next section, we actually display the exact solution $R_n$ in a very compact form.

An important remark is in order.
The relation \recurn\ is strikingly reminiscent of that for the orthogonal polynomials 
\mainequpsix\ say with $g_2=g_6=0$, $g_4=g$, except that the l.h.s. of \mainequpsix\
is now replaced by $1$, and $r_n=h_n/h_{n-1}$ by $R_n$. One may wonder whether eq.\recurn\
may be derived from some matrix model solution. The answer is not known to this day, but
the boundary condition that $R_{-1}=0$ would mean in matrix model language that some
norm of orthogonal polynomial must vanish, hence if there is such a matrix model formulation,
it must be very singular.
As to the r.h.s. of \mainequpsix, its similarity with that of \recurn\ suggests to express
the rules for the possible subtrees encountered around the vertex attached to the root
in {\it counterclockwise order}
in terms of a ``Q-operator" acting on a formal orthonormal basis $|n\rangle$, 
$\langle m|n\rangle=\delta_{m,n}$ for $m,n\geq 0$ and $|n\rangle =0$ for $n<0$, via
\eqn\qaction{ Q|n\rangle=|n+1\rangle + R_n |n-1\rangle }
The first term is interpreted as the contribution of a single black leaf, while the second
corresponds to a blossom tree with at most $n$  excess black leaves.
Then the r.h.s. of \recurn is nothing but $1+ g\langle n-1| Q^3|n\rangle$.

\subsec{Exact solution}

To solve \recurn, we use the convergence condition to write $R_n=R-\rho_n$ at large $n$,
and expand \recurn\ at first order in $\rho_n$. This gives the linear recursion relation
\eqn\linrec{ gR(\rho_{n+1}^{(1)}+\rho_{n-1}^{(1)})-\rho_n^{(1)} (1-4g R)=0}
This has the characteristic equation
\eqn\caraq{ x+{1\over x}+4={1\over gR} }
with $R$ given by \quareduce.
Picking the solution $x$ with modulus less than $1$,
we find that $R_n=R(1-\lambda_1 x^n +O(x^{2n}))$
for some integration constant $\lambda_1$. We may next expand $R_n=R(1-\sum_{j\geq 1} \lambda_j x^{jn})$,
and \recurn\ turns into a recursion relation for the coefficients $\lambda_j$:
\eqn\turninto{ \lambda_{j+1}\left(x^{j+1}+{1\over x^{j+1}}-x-{1\over x}\right)=\sum_{i=1}^j
\lambda_i\lambda_{j+1-i} \left(x^{i}+{1\over x^{i}}\right) }
solved recursively as 
\eqn\solarho{\lambda_j=\lambda_1\left({\lambda_1 x\over (1-x)(1-x^2)}\right)^{j-1}
{1-x^j\over 1-x}}
Picking $\lambda_1=x(1-x)(1-x^2)\lambda$, $R_n$ is easily resummed
into
\eqn\genericsol{ R_n=R {(1-\lambda x^{n+1})(1-\lambda x^{n+4})\over 
(1-\lambda x^{n+2})(1-\lambda x^{n+3})}}
Further imposing the initial condition $R_{-1}=0$ fixes $\lambda=1$, hence finally
\eqn\soltet{R_n=R{(1-x^{n+1})(1-x^{n+4})
\over (1-x^{n+2})(1-x^{n+3})} }
with $|x|<1$ solving \caraq.

This gives an explicit formula for the generating function of 4-valent two-leg graphs 
with geodesic distance at most $n$ between the legs. In particular, for $n=0$, this gives
the generating function for graphs with the two legs in the same face (also called
$\Gamma_2$ in Sect.3.3), namely
\eqn\rzero{ R_0=G_0=\Gamma_2=R{1+x^2\over 1+x+x^2}=R{(1-4gR)\over (1-3gR)}=R-gR^3 }
where we have used \caraq\ and \funcgenqua\ to simplify the result.
This is in perfect agreement with the matrix model result \quagamto.

\subsec{Integrability}

The equation \recurn\ is intergable in the classical sense that there exists an ``integral
of motion", namely a conserved quantity $f(R_n,R_{n+1})=$const. which implies \recurn.
More precisely, defining
\eqn\consqua{ f(x,y)=x y (1-g x-g y) -x-y }
we have
\eqn\quancons{ f(R_{n},R_{n+1})-f(R_{n-1},R_n)=(R_{n+1}-R_{n-1})
\big(R_n-1-gR_n(R_{n+1}+R_n+R_{n-1})\big)}

We deduce that if $f(R_n,R_{n+1})$is a constant independent of $n$, then $R_n$ obeys \recurn:
$f$ is an integral of motion of the equation \recurn. 

Using $f$, we may write in a compact way the condition $\lim_{n\to\infty}R_n=R$ for solutions of
\recurn. Indeed, we simply have to write
\eqn\frr{ f(R_n,R_{n+1})=f(R,R)=R^2(1-2gR)-2R=-(R-gR^3) }
All solutions to \frr\ are also solutions of \recurn, and they moreover converge to $R$ as $n\to\infty$.
As an immediate application of \frr, we may recover $R_0$, by imposing that $R_{-1}=0$:
\eqn\rzerf{ R_0=-f(R_{-1},R_0)=R-gR^3 }
in agreement with \rzero.

\subsec{Fractal dimension}

The advantage of having an exact formula like \soltet\ is that we may also extract the 
``fixed area" coefficient $R_{n,A}$ of $g^A$ in $R_n$ via the contour integral
\eqn\finrn{
R_{n,A}= \oint {dg \over 2i\pi g^{A+1}} R_n }
with $R_n$ given by \soltet.
This gives
access to asymptotic properties at large area $A$. In particular, the ratio
\eqn\behanum{ B_n\equiv \lim_{A\to \infty} {R_{n,A}\over R_{0,A}} } 
may be taken as a good estimate of the average number of points 
at a geodesic distance less or equal to $n$ from
a given point in random 4-valent graphs of infinite area. 
It is expected to behave like 
\eqn\behas{ B_n \sim n^{d_F} \ \ {\rm for} \ {\rm large}\ n }
where $d_F$ is the fractal dimension of the random surfaces.
Performing in \finrn\ the change of variables $v=gR$, i.e. $g=v(1-3v)$,
we obtain
\eqn\rnaob{ R_{n,A}=\oint {dv (1-6v) \over 2i\pi (v(1-3v))^{A+1}} 
{1\over 1-3v} {(1-x(v)^{n+1}) (1-x(v)^{n+4}) \over 
(1-x(v)^{n+2})(1-x(v)^{n+3})} } 
where we have used $R(g(v))=1/(1-3v)$ and the expression 
$x=x(v)\equiv (1-4v-\sqrt{1-8v+12 v^2})/(2v)$.
The large $A$ behavior is obtained by
a saddle-point approximation, as the integral is dominated by the vicinity 
of $v=v_c=1/6$, corresponding to the critical point $g=g_c=1/12$, where $x\to 1$. 
Making the change of variables $v=v_c(1+i{\xi\over \sqrt{A}})$, expanding
all terms in powers of $1/\sqrt{A}$ and integrating over $\xi$,
we finally get the leading behavior
\eqn\RnN{ R_{n,A} \sim {\rm const.}{(12)^A \over A^{5\over 2}} 
{(n+1)(n+4)\over (n+2)(n+3)} (140+270 n+179 n^2+50 n^3+5 n^4) }
which finally gives the ratio 
\eqn\bnexpr{ B_n = {3\over 280} {(n+1)(n+4)\over (n+2)(n+3)} 
(140+270 n+179 n^2+50 n^3+5 n^4)
\sim {3\over 56} n^4 }
hence $d_F=4$ is the desired fractal dimension.

\subsec{Scaling limit: Painlev\'e again!}

A continuum limit may be reached by letting $g$ tend
to its critical value $g_c=1/12$. 
More precisely, we write
\eqn\wriscaqua{ g={1\over 12}(1-\epsilon^4) \qquad \Rightarrow \qquad gR={1\over 6}(1-\epsilon^2)}
{}from eq.\quareduce. In turn, the characteristic equation \caraq\ yields
\eqn\xrt{ x=e^{-a\epsilon}+O(\epsilon^3) \qquad a=\sqrt{6} }
As seen from eq.\soltet, a sensible limit is obtained by writing
\eqn\scaling{n={r\over \epsilon}}
and letting $\epsilon\to 0$. Writing the scaling variable $r$ as $r=n/\xi$,
we see that $\epsilon$ plays the role of the inverse of the
correlation length $\xi$. 
As we approach the critical point, we have 
$\xi=\epsilon^{-1}= \big((g_c-g)/g_c\big)^{-\nu}$ with a critical exponent  
$\nu=1/4$, in
agreement with $\nu=1/d_F$, as expected from general principles. 
Performing this limit explicitly on the solution \soltet\ yields
an explicit formula for the continuum partition function ${\cal F}(r)$ of
surfaces with two marked points
at a geodesic distance {\it larger or equal to} $r$:
\eqn\exprneps{ {\cal F}(r)\equiv \lim_{\epsilon\to 0}{R-R_n\over \epsilon^2 R}=
-2 {d^2 \over dr^2} {\rm Log}\, \sinh \left({\scriptstyle\sqrt{3\over 2}}\, r\right)={3
\over \sinh^2\left({\scriptstyle\sqrt{3\over 2}}\, r\right)} }
Upon differentiating w.r.t. $r$, we obtain the continuum partition function
for surfaces with two marked points at a
geodesic distance {\it equal to} $r$: 
\eqn\exprderiv{{\cal G}(r)=-{\cal F}'(r)=
{3\sqrt{6}}\, {\cosh \left({\scriptstyle\sqrt{3\over 2}}\, r\right)\over \sinh^3
\left({\scriptstyle\sqrt{3\over 2}}\, r\right)}}
This reproduces a conjecture \AW\ obtained in a transfer matrix formalism 
of 2D quantum gravity.

Note that the precise form of the scaling function ${\cal F}(r)$ may alternatively
be obtained by solving the continuum counterpart of eq.\recurn. Indeed, 
writing 
\eqn\scaleq{R_n=R(1-\epsilon^2 {\cal F}(n\epsilon))}
and expanding eq.\recurn\ up to order $4$ in $\epsilon$, we obtain the
following differential equation
\eqn\painleveone{{\cal F}''(r)-3{\cal F}^2(r)-6{\cal F}(r)=0}
It is easy to check that ${\cal F}(r)$ as given by \exprneps\ is the 
unique solution of \painleveone\ with boundary conditions ${\cal F}(r)\to \infty$
when $r\to 0$ and ${\cal F}(r)\to 0$ when $r\to \infty$.
Writing ${\cal F}(r)=u(r)-1$, we note that eq.\painleveone\ turns into
\eqn\weier{u^2-u''/3=1} 
strikingly reminiscent of the Painlev\'e I equation governing
the model's all-genus double-scaling limit \pleve, except for the r.h.s. which is now
a constant. The function $u$ leading to $\cal F$ is simply the unique solution to
\weier\ such that $u(0^+)=\infty$ and $u(+\infty)=1$.

\subsec{Generalizations}

The results of Sects.6.1-6.5 generalize straightforwardly to the case of arbitrary even valences.
Using again the bijection of Sect.4.2, we still have to keep track of excess black leaves.
Introducing similarly the generating function $R_n$ for planar graphs with even valences and with
two legs at geodesic distance less or equal to $n$, we get a recursion relation by inspecting
all configurations of the vertex attached to the root of the corresponding blossom trees.
We may use the same rules as those found in the 4-valent case \recurn.
Going clockwise around the vertex and starting from the root,
we may encounter blossom trees with up to $p$ excess black leaves or single black leaves.
Encountering a black leaf decreases
the index $p$ of the objects following it clockwise, while encountering a blossom subtree
increases it by $1$. Using the ``Q-operator" formalism of Sect.6.1, namely that
$Q |n\rangle =|n+1\rangle+R_n |n-1\rangle$, we get the 
general recursion relation
\eqn\genercun{ R_n=1+\sum_{k\geq 1} g_{2k} \langle n-1| Q^{2k-1} | n\rangle }
to be supplemented with $d/2-1$ initial conditions $R_{-1}=R_{-2}=...R_{d/2-1}=0$ ($d={\rm deg}(V)$),
and the usual convergence condition $\lim_{n\to \infty} R_n=R$, to the solution $R$ of \Reveneq.
The explicit solution to \genercun\ with these boundary conditions was derived in \GEOD, and involves
soliton-like expressions. It allows for investigating the fractal dimension for multicritical
planar graphs, found to be $d_F=2(m+1)$ for the case of Sect.3.6 \talom, and to derive continuum
scaling functions for multicritical matter on surfaces with two marked points at a fixed geodesic 
distance $r$. 
Writing \genercun\ as $1=\langle n-1|V'(Q)|n\rangle$, we use again the trick of adding a weight $t$ 
per face of the graph, which amounts to replacing $V'(Q)\to V'(tQ)$, and multiplying by $t$
leaves us with $t=\varphi(t^2R_n,t^2R_{n\pm 1},...)$. Taking the multicritical values for $g_{2k}$,
and writing $t=t_c(1-\epsilon^{2(m+1)})$,
we look for solutions of \genercun\
of the form $R_n=R(1-\epsilon^2 {\cal F}(r=n\epsilon))$. This gives at order $2(m+1)$ in $\epsilon$
a differential equation for $\cal F$. Noting that our scaling Ansatz for $R_n$ is the same as that
for the double-scaling limit ($r_n=r_c(1-a u(y))$) except for the prefactor $R=R_c(1-\epsilon^2)$ 
we see that $u(r)=1+{\cal F}(r)$ satisfies the generalized Painlev\'e equation \redsimp,
but with a constant r.h.s. In differentiated form, this corresponds to writing
the commutation relation $[P,Q]=0$ between two differential operators $P$ and $Q$ of the variable $r$,
with respective orders $2m+1$ and $2$, with $Q=d^2-u$.

The generalization to graphs with arbitrary (even and odd) valences is straightforward, as we simply
have to use the ``Q-operator" formalism in the combinatorial setting. The functions $S_n$ (resp. $R_n$)
generate planar graphs with one leg (resp. two legs), with the leg (resp. second leg) at distance
at most $n$ from the external face.
The operator $Q$ now acts as $Q|n\rangle =|n+1\rangle +S_n|n\rangle +R_n|n-1\rangle$, where the 
new contribution corresponds to subtrees of charge $0$, that do not affect the numbers of allowed excess
black leaves of their followers. We obtain the system of equations
\eqn\systpla{ 0=\langle n|V'(Q)|n\rangle \qquad  1=\langle n-1|V'(Q)|n\rangle }

This generalizes presumably to all planar graph enumeration problems for which a matrix model
treatment is available, using orthogonal polynomials involving a natural $Q$ operator, 
interpreted in the combinatorial setting as describing objects of various charges attached
to the root vertex of the corresponding blossom trees.
We may infer that in the general multicritical case of a CFT with central charge $c(p,q)<1$,
the scaling function for surfaces with two marked points at geodesic distance at least $r$
is governed by a differential system of the form $[P,Q]=0$, $P$ and $Q$ two differential operators
of the variable $r$ of respective degrees $p$ and $q$.

\newsec{Planar graphs as spatial branching processes}

This last section is devoted to a dual approach to that followed so far, in which we
consider the graphs dual to those contributing to the matrix model free energy,
namely with prescribed face valences rather than vertex valences. On such a graph,
the geodesic distance between vertices is the minimal number of edges visited in a path
from one to the other. We will present bijections between classes of such graphs with
a specified origin vertex and with a marked vertex at geodesic distance $\leq n$, 
and labeled trees of arbitrary valences obeying some specific labeling rules.

This allows to make the contact with an active field of probability theory dealing with
spatially branching processes. The following is largely based on refs. \CS\ \ONEWALL\ 
\TWOWALL\ \MOB.

We first concentrate on the quadrangulations, namely the duals of 4-valent graphs.

\subsec{The dual bijections: labeled trees for planar quadrangulations}

\fig{The bijection betwen planar quadrangulations and labeled trees.
A planar rooted quadrangulation (a) and the natural labeling of its vertices by the 
geodesic distance to the origin vertex of the rooted edge (arrow). The confluent faces
are shaded. The tree edges are represented in thick black lines, and connect
all vertices with positive labels. Erasing all but these new edges and the vertices
they connect leaves us with a labeled tree (b), which we root at the vertex
corresponding to the end of the rooted edge of the initial quadrangulation.
Finally, all labels of the tree are shifted by $-1$.}{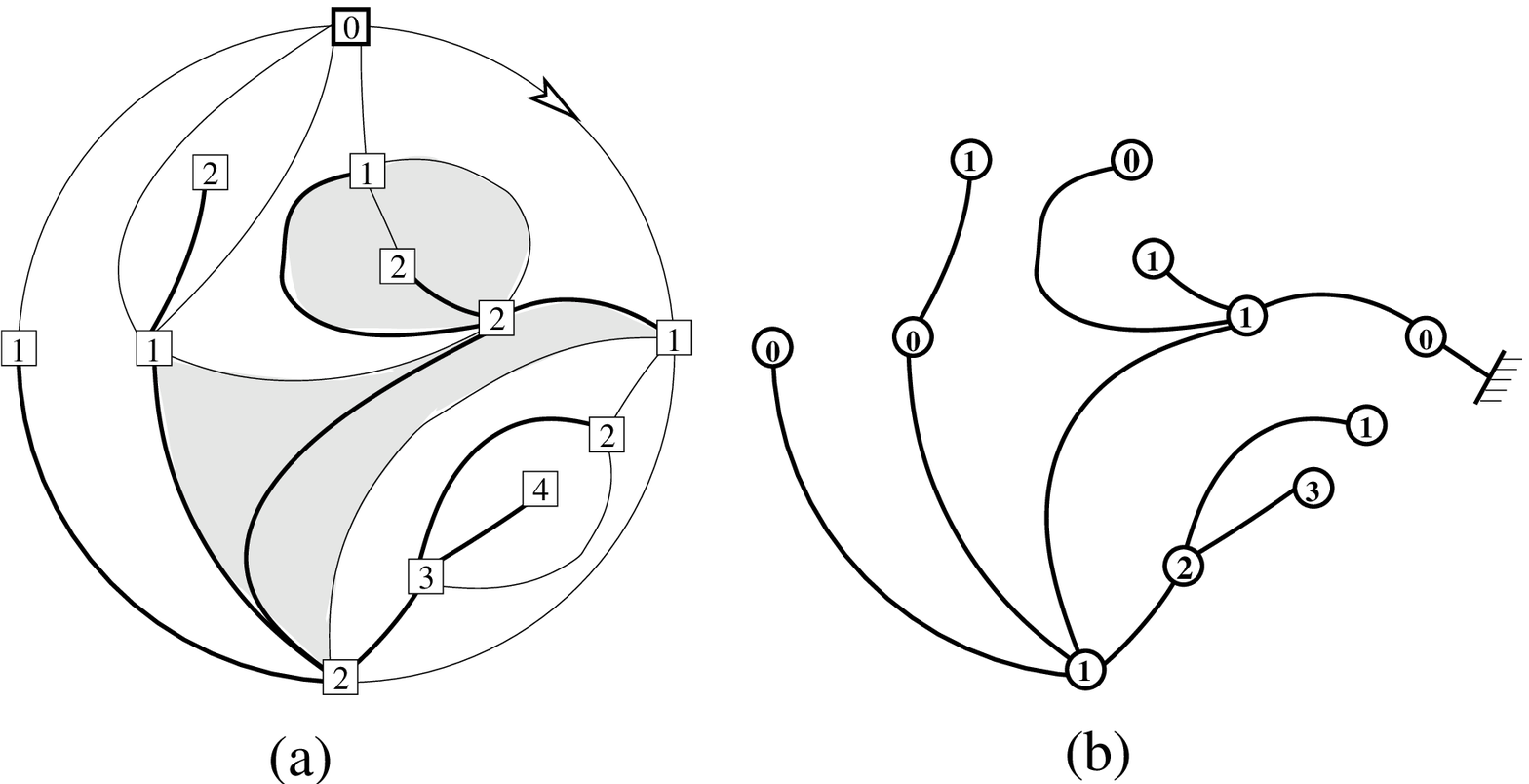}{12.cm}
\figlabel\quadtree

We start with a rooted planar quadrangulation, namely a graph with only 4-valent faces
(squares), with a marked oriented ``root" edge. Let us pick as origin vertex the vertex 
at which the root edge starts. This choice induces a natural labeling of the vertices
of the graph by their geodesic distance to this origin, itself labeled $0$ (see Fig.\quadtree\ (a)
for an example).
We then note that only two situations may occur for the labeling of vertices around a face,
namely
\eqn\twositu{ \figbox{8.cm}{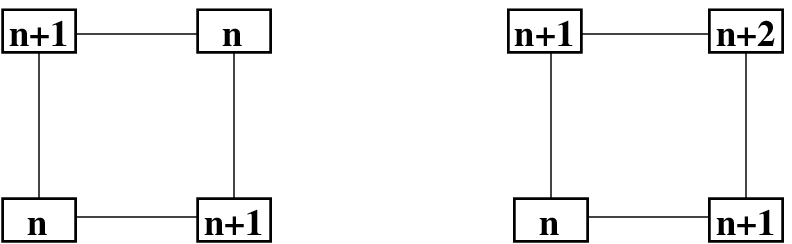} }
in which cases the faces are respectively called confluent and normal. The confluent faces have been shaded in
the example of Fig.\quadtree\ (a). We now construct new edges
as follows:
\eqn\twotree{ \figbox{8.cm}{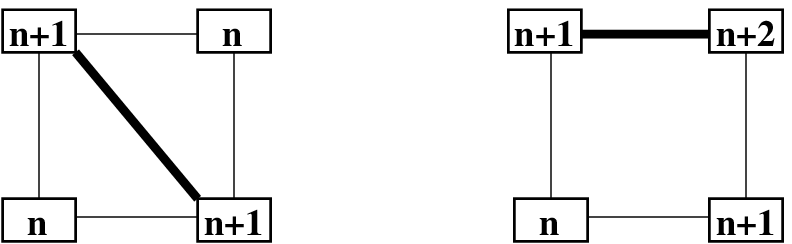} }
in each face of the quadrangulation (including the external face, for which the rules are reversed).
This rule may be summarized by saying that we connect via a new edge all the vertices
immediately followed clockwise by a vertex with a label one less.
These edges are readily seen to connect all vertices of the quadrangulation but the origin.
Thus, erasing all but the new edges and the vertices they connect leaves
us with a connected labeled tree (see Fig.\quadtree\ (b)), which we root at the end vertex 
of the original rooted edge
of the quadrangulation, and in which we {\it subtract $1$ from all vertices}\foot{This is just a 
technical trick to make the precise contact with the generating function $R_n$ of Sects.4 and 6.
The reader will have to remember to add up one to each vertex label of the tree to recover
its geodesic distance from  the origin in the quadrangulation.}. 
In particular, the vertex attached to the root has label $0$, and all labels are non-negative.
Moreover, by the construction rules \twotree, adjacent labels of the tree may differ 
only by $0$ or $\pm 1$. Such trees are called well-labeled, and are in bijection with the
rooted planar quadrandulations.

The construction rules \twotree\ allow for interpreting the features of the tree in terms of the original
quadrangulation. Any vertex labeled $n-1$ in the tree corresponds to a vertex at distance $n$ from the origin
in the quadrangulation. From the rules of eq.\twotree,
we see that any marked edge $n\to n+1$ of the quadrangulation corresponds 
marking an edge of the tree adjacent to a vertex labeled $n$. This in turn may be viewed as the rooting
of the tree at a vertex labeled $n$ (the above bijection uses this fact for $n=0$).

We next define rooted well-labeled trees as rooted labeled trees, with non-negative integer
vertex labels, and such that the root vertex has label $n$. Let $R_n$ be the generating 
function for such objects, with a weight $g$ per edge. According to the above bijection,
the generating function for rooted planar quadrangulations with a weight $g$ per face is simply $R_0$. 
If, instead of rooting the well-labeled tree at the end vertex of the initial quadrangulation,
we had chosen to root it elsewhere, typically at another vertex of the tree say labeled $n$,
the resulting rooted well-labeled tree would satisfy the extra condition that the label
$0$ occurs at least once in the tree. The generating function for such an object is nothing
but $G_n=R_n-R_{n-1}$. In terms of the original quadrangulation, this is nothing but
the generating function of quadrangulations with an origin vertex and with a marked edge $n\to n+1$
w.r.t. this origin.
So $R_n$ is the generating function for planar quadrangulations with an origin
and with a marked edge $m\to m+1$, $m\leq n$, and a weight $g$ per face.

The definition of $R_n$ allows to derive a recursion relation of the form
\eqn\recunewrn{ R_n={1\over 1-g (R_{n+1}+R_n+R_{n-1})} }
where we simply express the labeling rule that the root vertex labeled $n$ may be adjacent
to any number of vertices labeled $n,n+1$ or $n-1$, themselves roots of other well-labeled trees.
Moreover, for \recunewrn\ to also make sense at $n=0$ we must set $R_{-1}=0$. 
Removing the constraint that $m\leq n$ by sending $n\to \infty$ leaves us with the generating
function $R$ for quadrangulations with an origin and a marked edge, wich also generates
the rooted quadrangulations with a marked vertex, and should satisfy the relation
\eqn\relaR{ R={1\over 1-3 g R} }
with $R=1+O(g)$.  We conclude that the functions $R$ and $R_n$ coincide with those introduced in
Sects.4.1 and 6.1.

So we have found another (dual) combinatorial interpretation for the exact solutions \soltet.

\subsec{Application I: average numbers of edges and vertices at distance
$n$ from a vertex in quadrangulations}

A direct application of this new interpretation of $R_n$ concerns properties
of large random quadrangulations viewed from their origin. For instance, the average 
$\langle e_n\rangle_A$
of the number of edges $n\to n+1$ in a quadrangulation with an origin and with say $A$ faces
is given by
\eqn\avratio{ {\langle e_n\rangle \over \langle e_0 \rangle } = {R_{n,A}-R_{n-1,A}\over R_{0,A}}}
with $R_{n,A}$ as in \finrn. Again, this is readily 
computed in the limit $A\to \infty$, where we first note $\langle e_0 \rangle\to 4$ by 
Euler's relation, and then use a saddle point method just like in \RnN, resulting in
\eqn\calcen{\langle e_n\rangle={6\over 35} {
(n^2+4n+2)(5n^4+40n^3+117n^2+148n+70) \over (n+1)(n+2)(n+3) } }
This goes as $6n^3/7$ for large $n$, which confirms the value $d_H=4$ for the fractal dimension,
as $\langle e_n\rangle \sin d/dn \, n^{d_F} \sim n^{d_F-1}$.

We may also obtain the average number of vertices at geodesic distance $n$ from 
the origin, by noting that the corresponding generating function is that of unrooted
well-labeled trees with at least a label $0$ and a marked vertex with label $n-1$.
Abandoning the condition that a label $0$ should occur, and
decomposing the tree according to the environment of the marked vertex with label $n-1$
results in the generating function
\eqn\genkol{ K_{n-1}=\sum_{k=1}^\infty 
{g^k\over k} (R_{n}+R_{n-1}+R_{n-2})^k=
-{\rm Log}\big(1-g(R_{n}+R_{n-1}+R_{n-2})\big) =Log(R_{n-1})}
where we have incorporated the symmetry factor $1/k$ when the vertex has valence $k$. 
Finally, the generating function for quadrangulations with an origin and a marked vertex
at distance $n$ is
\eqn\bonnegf{ V_n=K_{n-1}-K_{n-2}={\rm Log}\left({R_{n-1}\over R_{n-2}}\right) }
for $n\geq 2$ and Log$R_0$ for $n=1$, while of course $V_0=1$.
Therefore the average number of vertices at distance $n$ from the origin 
in a quadrangulation of area $A$ is given by
\eqn\givn{ \langle v_n\rangle_A= {\rm Log}\left({R_{n-1,A}\over R_{n-2,A}}\right) }
easily derived in the large $A$ limit:
\eqn\givnAinf{\langle v_n\rangle={3\over35}
\big((n+1)(5n^2+10n+2)+\delta_{n,1}\big)} 
This goes as $3n^3/7$ for large $n$, also in agreement with $d_F=4$.

Note that eqs.\genkol\-\bonnegf\ also allow to interpret Log$R_{n-1}$ as the generating function
for quadrangulations with an origin and a marked vertex at distance $m\leq n$. In the limit
$n\to \infty$, the function Log$R$ therefore generates the quadrangulations with two
marked vertices. In the dual formulation, this corresponds to 4-valent planar graphs with two
marked faces: this gives a purely combinatorial derivation in the 4-valent case
of the formula \rewfintf\ obtained above in the matrix model language.

\subsec{Application II: local environment of a vertex in quadrangulations}

Another application of this new graph interpretation of $R_n$ concerns the local environment 
of the origin. Assume we wish to keep track of the numbers of vertices at some finite 
distances $p+1$ from the origin, and 
edges labeled $q\to q+1$ for some specific $p$'sand $q$'s, both less or equal to some given $k$. 
Then a way to do it
is to add extra weights, say $\rho_p$ per vertex labeled $p$ in the corresponding well-labeled tree
and $\sigma_p$ per edge adjacent to a vertex labeled $p$ of the well-labeled tree. 
Indeed, as explained in the previous section,
this amounts to adding a weight $\rho_p$ per vertex labeled $p+1$ in the quadrangulation, and a
weight $\sigma_p$ per edge $p\to p+1$ in the quadrangulation.
This turns the equation \recunewrn\ into a new set of equations 
\eqn\newset{ R_n={\rho_n \over 1- 
g \sigma_n R_n(\sigma_{n+1} R_{n+1}+\sigma_n R_n+\sigma_{n-1}R_{n-1})} , \qquad n=0,1,2,...,k+1}
with $\rho_{k+1}=\sigma_{k+1}=1$,
while $R_n$ satisfies \recunewrn\ for all $n\geq k+2$. This is slightly simplified by introducing 
$Z_n=\sigma_n R_n$ (with $\sigma_n=1$ for $n\geq k+1$), as we are left with
\eqn\eqszr{\eqalign{
Z_n&= {\sigma_n \rho_n\over 1- g \sigma_n(Z_{n+1}+Z_n+Z_{n-1})},\quad n=0,1,2,...,k\cr
Z_n&= {1\over 1- g (Z_{n+1}+Z_n+Z_{n-1})},\quad n=k+1,k+2,...\cr}}

Solving such a system seems quite difficult
in general, but we may use the integral of motion \consqua\ to replace the infinite set
of equations on the second line of \eqszr\ (and the convergence condition of $Z_n$ to $R$),
by simply the conserved quantity
\eqn\consk{ f(Z_{k},Z_{k+1})=f(R,R) }
Together with the first line of \eqszr, this gives a system of $k+2$ algebraic relations for the functions
$Z_0,Z_1,...,Z_{k+1}$, which completely determines them order by order in $g$.
As an example, let us compute in the case $k=0$
the generating function including a weight $\rho_0=\rho$ per vertex labeled
$0$ in the trees and $\sigma_0=\sigma$ per edge incident to a vertex labeled $0$
in the trees. (This in turn corresponds in the 
quadrangulations to a weight $\rho$ per vertex labeled $1$, i.e. per nearest neighbor of the origin, 
and a weight $\sigma$ per edge $0\to 1$.)
We get the system:
\eqn\rezerone{ Z_0={\rho\sigma \over 1-g\sigma (Z_0+Z_1)},\qquad Z_0Z_1(1-g(Z_0+Z_1)-Z_0-Z_1=f(R,R)=gR^3-R}
which upon eliminating $Z_1$ and reinstating $R_0=Z_0/\sigma$, boils down to
\eqn\boildown{ (R_0-\rho)(1+R_0-g\sigma^2 R_0^2-\rho) 
-\sigma R_0(R_0-\rho+gR(1-gR^2))+g\sigma^3R_0^3=0 }
for the generating function $R_0$ for rooted quadrangulations with weights $\rho$ per neighboring
vertex of the origin and $\sigma$ per edge adjacent to the origin.
$R_0\equiv R_0(g\vert \rho,\sigma)$ is the unique solution to \boildown\ such that $R_0=\rho+O(g)$.
Note that we recover $R_0=R-gR^3$ of \rzerf\ when $\rho=\sigma=1$.
As the rooting of the quadrangulation is itself a choice of an edge adjacent to the origin, we may 
express the corresponding generating function for ``unrooted " quadrangulations, namely with just
an origin vertex, as
\eqn\genoroot{ \Gamma_0(g\vert \rho,\sigma) =\int_0^\sigma {ds \over s}  R_0(g\vert \rho,s) }
simply expressing the rooting of the quadrangulation as $\sigma\partial_\sigma \Gamma_0=R_0$. 
The statistical average over quadrangulations of area $A$ of $\rho^{N_1}\sigma^{N_{01}}$
($N_1$ the number of neighboring vertices of the origin, $N_{01}$ the number edges adjacent
to the origin) finally reads
\eqn\finstat{ \langle \rho^{N_1}\sigma^{N_{01}} \rangle_A= {\Gamma_{0,A}(\rho,\sigma)
\over \Gamma_{0,A}(1,1) }= {\int_0^\sigma {ds \over s} R_{0,A}(\rho,s) \over 
\int_0^1 {ds \over s}  R_{0,A}(1,s)}}
where as usual $\Gamma_{0,A}(\rho,\sigma)$ (resp. $R_{0,A}(\rho,s)$) denotes the coefficient of $g^A$ in 
$\Gamma_{0}(g\vert \rho,\sigma)$ (resp. $R_0(g\vert \rho,s)$). The limit $\lim_{A\to\infty}
\langle \rho^{N_1}\sigma^{N_{01}} \rangle_A=\Gamma$ may again be extracted by a 
saddle-point expansion. After some algebra, we find
\eqn\afteralg{ 6 \Gamma(\Gamma+1)(\Gamma+3)-\sigma \big(2\Gamma(1+4\Gamma+\Gamma^2)+3\rho(\Gamma+1)^2
(\Gamma+2)\big)=0}
and $\Gamma$ is uniquely determined by the condition $\Gamma=1$ for $\sigma=\rho=1$.
For instance, when $\sigma=1$, we get 
\eqn\gamarho{ \Gamma(\rho,1)={2\over \sqrt{4-3\rho}}-1=\sum_{n\geq 1} \rho^n\left({3\over 16}\right)^n
{2n\choose n} }
in which we read the probability $P(n)=(3/16)^n{2n\choose n}$
for a vertex to have $n$ neighboring vertices in an infinite quadrangulation.
Similarly, taking $\rho=1$, we get
\eqn\simieq{  \Gamma(1,\sigma)={1\over 2}\left(\sqrt{6+3\sigma\over 6-5\sigma} -1\right) }
which generates the probabilities to have $n$ edges adjacent to a vertex in an infinite quadrangulation.
We may also derive the generating function for the conditional probabilities of having $n$
nearest neighboring vertices, given that there is no multiple edge connecting them to the origin,
by simply taking $\Gamma(\rho=t/\sigma,\sigma)$ and letting $\sigma\to 0$, which indeed suppresses all
contributions from multiply connected vertices. This gives
\eqn\mpd{ \Pi(t)=\lim_{\sigma\to 0}\Gamma\left({t\over \sigma},\sigma\right)= \sqrt{8-t\over 2-t} -2}
For instance, the probability that a given vertex have no multiple neighbors in an infinite 
quadrangulation is
\eqn\nomun{ \Pi(1)=\sqrt{7}-2 }

\subsec{Spatial branching processes}

We have seen so far how the information on the geodesic distance from the origin in a 
rooted planar quadrangulation
may be coded by rooted well-labeled trees. The latter give rise to natural examples of so-called 
spatially branching processes, in the context of which quantities like $R_n$ correspond to 
certain probabilities.

A spatial branching process consists of two data. First we have a monoparental
population, whose genealogy is described by a rooted tree, the root corresponding to the common
ancestor. A standard measure on these trees attaches
the probability $(1-p)p^k$ for any vertex to have $k$ descendents. The second data
is a labeling of the vertices of the tree
by positions say on the integer line $n\in \IZ$. Here, we add the rule of the ``possessive
ancestor" that his children must be at close enough positions from his 
(namely differing by $0$ or $\pm 1$).
Let $E(T)$ denote the probability of extinction of the population at generation $T$, then we
have the recursion relation
\eqn\recuextinct{ E_n(T)={1-p\over 1-{p\over 3}(E_{n+1}(T-1)+E_n(T-1)+E_{n-1}(T-1)) }}
Letting $T\to \infty$, we see that the extinction probability $E_n=\lim_{T\to \infty} E_n(T)$
obeys the same equation as $R_n$ \recunewrn\ upon some rescaling, and we find that
$E_n=(1-p)R_n\left(g={p(1-p)\over 3}\right)$, in the case of  positions restricted to
lie in a half-line (with a ``wall" at the origin). Without this restriction, the problem becomes
translationally invariant and $E_n=E=(1-p)R\left(g={p(1-p)\over 3}\right)$. Note that the critical
point $g=g_c=1/12$ corresponds here to the critical probability $p=p_c=1/2$.

In this new setting, we may ask different questions, such as what is the probability for the
process to escape from a given interval, say $[0,L]$. Once translated back into $R_n$ terms,
this amounts to still imposing the recursion relation \recunewrn, but changing boundary
conditions into
\eqn\chbound{ R_{-1}=0 \qquad {\rm and} \qquad R_{L+1}=0}
The escape probability from the interval reads then
\eqn\escaprob{ S_n=1-(1-p)R_n\left(g={p(1-p)\over 3}\right)=(1-p)(R-R_n) }
The equation \recunewrn\ with the boundary conditions \chbound\ still 
admits an exact solution expressed by means of the Jacobi $\theta_1$ function
\eqn\jacob{ \theta_1(z)=2i\sin(\pi z)\prod_{j\geq 1}
(1-2 q^j \cos(2\pi z)+q^{2 j})}
The solution reads $R_n=R_n^{(L)}$, with
\eqn\solutet{\eqalign{
R_n^{(L)}&=R {u_n u_{n+3}\over u_{n+1} u_{n+2} }\cr 
u_n&=\theta_1\left({n+1\over L+5}\right) \cr}}
guaranteeing that the boundary conditions \chbound\ are satisfied, and
where the nome $q$ still has to be fixed.
The main recursion relation \recunewrn\ reduces to a quartic
equation for the $u_n$'s:
\eqn\qartiu{
u_nu_{n+1}u_{n+2}u_{n+3}= {1\over R} u_{n+1}^2u_{n+2}^2+g R(u_{n-1}u_{n+2}^2u_{n+3}+
u_n^2u_{n+3}^2+u_n u_{n+1}^2u_{n+4})}
and the latter is satisfied by \solutet\ provided we take
\eqn\wetak{\eqalign{ R&= 4 {\theta_1(\alpha) \theta_1(2\alpha)\over
\theta_1'(0)\theta_1(3\alpha)} \left({\theta_1'(\alpha)\over \theta_1(\alpha)}-{1\over 2}
{\theta_1'(2\alpha)\over \theta_1(2\alpha)}\right)\cr
g&={\theta_1'(0)^2\theta_1(3\alpha)\over 16\theta_1(\alpha)^2\theta_1(2\alpha) 
\left({\theta_1'(\alpha)\over \theta_1(\alpha)}-{1\over 2}
{\theta_1'(2\alpha)\over \theta_1(2\alpha)}\right)^2}\cr} }
for $\alpha=1/(L+5)$. The identity \qartiu\ is proved typically by showing that both sides
have the same transformations under $n\to n+L+5$ and $n\to n+(L+5)/(2i\pi){\rm Log}q$, and that
moreover they have the same zeros, this latter condition amounting to \wetak.

The elliptic solution $R_n$ may be interpreted terms of {\it bounded} graphs as follows. 
The quantity $G_n^{(L)}=R_n^{(L)}-R_{n-1}^{(L-1)}$ is the generating function
for quadrangulations with an origin and a marked edge $n\to n+1$, which are moreover
bounded in the sense that all vertices are distant by at most $L+1$ from the origin.

Taking again the continuum scaling limit of the model
leads to the probabilists' Integrated SuperBrownian Excursions (ISE), 
here in one dimension \ISE. The scaling function $\cal U$ obtained from 
$R_n=R_c(1-\epsilon^2{\cal U})$ in the limit \wriscaqua, while moreover $r=n\epsilon$ and
$\lambda=(L+5)\epsilon$ are kept fixed, reads:
\eqn\neubp{{\cal U}(r)=2\wp(z|\omega,\omega') }
where $\wp$ is the Weierstrass
function ($\wp=-\partial_r^2{\rm Log}\,\theta_1$),
with half-periods $\omega=\lambda/2$ and $\omega'$, related via the condition
that the second invariant $g_2(\omega,\omega')=3$.

\subsec{Generalizations}

We have so far only discussed quadrangulations and their relations to spatial branching
processes (see also \CHASDUR\ \MARMO). 
All of the above generalizes to rooted
planar graphs with arbitrary even face valences. 
These are in bijection with rooted well-labeled trees with
more involved labeling rules, also called well-labeled mobiles \MOB. 
This allows for a generalization of spatial branching processes, possessing
these labeling rules. As we already know that these objects have an interesting 
variety of multicritical behaviors, this should turn into multicritical generalizations 
of the ISE.

In \MOB, the general case covered by two-matrix models is treated as well, and seen
to generate Eulerian (i.e. vertex-bicolored) planar graphs. 
The latter contain as a particular case the gravitational Ising model, 
and in principle allow for reaching any $c(p,q)$ CFT coupled to 2D quantum gravity.
These will lead presumably to interesting generalizations of the ISE.
 
\newsec{Conclusion}

In these lectures we have tried to cover various aspects of discrete 2D quantum
gravity, namely of statistical matter models defined on random graphs of given
topology. 

The matrix model approach, when solvable, gives exact recursion relations
between quantities eventually leading to compact expressions for the genus expansion
of the free energy of the models. We have further investigated the so-called double scaling
limit in which both matter and space degrees of freedom become critical, allowing
for instance to define and compute a scaling function summarizing the leading singularities
of the free energy at all genera, as a function of the renormalized cosmological constant
$x$. 
The final general result takes the form
\eqn\canonagain{ [P,Q]=1,\qquad P=d^p+v_2 d^{p-2}+...+v_p,\quad Q=d^q+u_2d^{q-2}+...+u_q}
with $d=d/dx$, all $v$'s and $u$'s functions of $x$, and $u_2$ proportional to $F''$, the 
second derivative of the singular part of
the all-genus free energy w.r.t $x$.

The combinatorial approach, when bijections with trees are available, also gives exact
recursion relations between basic generating functions which can be interpreted in
terms of planar graph counting, while keeping track of the geodesic distance
between marked points. The expressions for the solutions are completely explicit,
allowing for taking a scaling limit, describing the free energy for
random surfaces with marked
points at a renormalized geodesic distance $r$. We may write the general result
for this scaling free energy in the form
\eqn\againcano{[P,Q]=0,\qquad P=d^p+v_2 d^{p-2}+...+v_p,\quad Q=d^q+u_2d^{q-2}+...+u_q}
with $d=d/dr$, all $u$'s and $v$'s functions of $r$, and $u_2$ proportional to the scaling
two-point function for surfaces with two marked points at geodesic distance $\geq r$.

Remarkably, in all cases solved so far, the exactly solvable geodesic distance problems
for planar graphs all correspond to cases where a matrix model solvable by orthogonal polynomials
is available. It seems therefore that the bijections with trees exactly parallel the
orthogonal polynomial solutions. More precisely, we have observed that a similar
abstract ``Q-operator" could be introduced in both cases, one of them describing the
possible subtrees one can encounter when going counterclockwise around a vertex of a 
blossom tree, the other describing the multiplication by an eigenvalue $\lambda$
on the basis of orthogonal polynomials.

The two apparently unrelated results \canonagain\ and \againcano\
show that something deeper happens here, that
deserves to be better understood. One may imagine that there must exist a more general structure
which would unify and combine the notions of genus and geodesic distance, and give
for instance closed equations for scaling functions of both $x$ and $r$. To reach this,
one should first be able to control geodesic distances in higher genus as well, by
generalizing the tree bijection techniques explained here only in the planar case. 
Another possibility could be that matrix models as we know them today may still be
only part of a more general setting. Some generalizations of matrix models involving
integration of eigenvalues over contours (or linear combinations thereof) in the
complex plane may be the correct answer, and relate to the intrinsic
geometry of graphs once interpreted combinatorially.

Finally, it is interesting to notice that no continuum field theoretical representation
of geodesic distance dependence of random surfaces has been found yet, although 2D quantum
gravity is now well understood in terms of the coupling of CFT to the Liouville field theory
\KPZ. The simplicity of the results found here for the various scaling functions comes as a
surprise in  that respect. Field theory probably still has some way to go before explaining the
purely combinatorial results shown here.

\listrefs
\end

%% file: toc.tex
\noindent {1.} {Introduction} \leaderfill{2} \par 
\noindent \quad{1.1.} {Matrix models {\fam \itfam \tenit per se}} \leaderfill{2} \par 
\noindent \quad{1.2.} {A brief history} \leaderfill{3} \par 
\noindent {2.} {Matrix models for 2D quantum gravity} \leaderfill{4} \par 
\noindent \quad{2.1.} {Discrete 2D quantum gravity} \leaderfill{4} \par 
\noindent \quad{2.2.} {Gaussian integral's diagrammatics} \leaderfill{4} \par 
\noindent \quad{2.3.} {Gaussian matrix integral and more diagrammatics} \leaderfill{6} \par 
\noindent \quad{2.4.} {Model building I: using one-matrix integrals} \leaderfill{9} \par 
\noindent \quad{2.5.} {Model building II: using multi-matrix integrals} \leaderfill{12} \par 
\noindent {3.} {The one-matrix model I: large $N$ limit and the enumeration of planar graphs} \leaderfill{14} \par 
\noindent \quad{3.1.} {Eigenvalue reduction} \leaderfill{14} \par 
\noindent \quad{3.2.} {Large size: the saddle-point technique} \leaderfill{15} \par 
\noindent \quad{3.3.} {Enumeration of planar graphs with external legs} \leaderfill{19} \par 
\noindent \quad{3.4.} {The case of 4-valent planar graphs} \leaderfill{22} \par 
\noindent {4.} {The trees behind the graphs} \leaderfill{23} \par 
\noindent \quad{4.1.} {4-valent planar graphs and blossom trees} \leaderfill{23} \par 
\noindent \quad{4.2.} {Generalizations} \leaderfill{26} \par 
\noindent {5.} {The one-matrix model II: topological expansions and quantum gravity} \leaderfill{27} \par 
\noindent \quad{5.1.} {Orthogonal polynomials} \leaderfill{27} \par 
\noindent \quad{5.2.} {Large $N$ limit revisited} \leaderfill{30} \par 
\noindent \quad{5.3.} {Singularity structure and critical behavior} \leaderfill{31} \par 
\noindent \quad{5.4.} {Higher genus} \leaderfill{32} \par 
\noindent \quad{5.5.} {Double-scaling limit} \leaderfill{33} \par 
\noindent \quad{5.6.} {Generalization to multi-matrix models} \leaderfill{37} \par 
\noindent {6.} {The combinatorics beyond matrix models: geodesic distance in planar graphs} \leaderfill{38} \par 
\noindent \quad{6.1.} {Keeping track of the geodesic distance: the 4-valent case} \leaderfill{39} \par 
\noindent \quad{6.2.} {Exact solution} \leaderfill{40} \par 
\noindent \quad{6.3.} {Integrability} \leaderfill{41} \par 
\noindent \quad{6.4.} {Fractal dimension} \leaderfill{42} \par 
\noindent \quad{6.5.} {Scaling limit: Painlev{\accent 19 e} again!} \leaderfill{43} \par 
\noindent \quad{6.6.} {Generalizations} \leaderfill{44} \par 
\noindent {7.} {Planar graphs as spatial branching processes} \leaderfill{45} \par 
\noindent \quad{7.1.} {The dual bijections: labeled trees for planar quadrangulations} \leaderfill{46} \par 
\noindent \quad{7.2.} {Application I: average numbers of edges and vertices at distance $n$ from a vertex in quadrangulations} \leaderfill{48} \par 
\noindent \quad{7.3.} {Application II: local environment of a vertex in quadrangulations} \leaderfill{50} \par 
\noindent \quad{7.4.} {Spatial branching processes} \leaderfill{52} \par 
\noindent \quad{7.5.} {Generalizations} \leaderfill{54} \par 
\noindent {8.} {Conclusion} \leaderfill{54} \par